\documentclass{article}

\usepackage{color}
\usepackage{amsmath, amssymb}

\newcommand{\RM}{\mathbb{R}}
\newcommand{\ZM}{\mathbb{Z}}

\newcommand{\CM}{\mathbb{C}}
\newtheorem{theorem}{Theorem}
\newtheorem{lemma}{Lemma} 
\newtheorem{prop}{Proposition} 
\newtheorem{cor}{Corollary}

\newcommand{\xvec}{\ensuremath{\boldsymbol{x}}}

\newcommand{\evec}{\ensuremath{\boldsymbol{e}}}

\newcommand{\kvec}{\ensuremath{\boldsymbol{k}}}
\newcommand{\wvec}{\ensuremath{\boldsymbol{w}}}

\begin{document}

\title{{\bf Walk/Zeta Correspondence}
\vspace{15mm}}

\author{Takashi KOMATSU \\
Department of Bioengineering, School of Engineering \\ 
The University of Tokyo \\
Bunkyo, Tokyo, 113-8656, JAPAN \\ 
e-mail: komatsu@coi.t.u-tokyo.ac.jp 
\\ \\ 
Norio KONNO \\
Department of Applied Mathematics, Faculty of Engineering \\ 
Yokohama National University \\
Hodogaya, Yokohama, 240-8501, JAPAN \\
e-mail: konno-norio-bt@ynu.ac.jp \\
\\ 
Iwao SATO \\ 
Oyama National College of Technology \\
Oyama, Tochigi, 323-0806, JAPAN \\ 
e-mail: isato@oyama-ct.ac.jp 
}

\date{\empty }

\maketitle

\vspace{50mm}


\vspace{20mm}











\clearpage

\begin{abstract}
Our previous work presented explicit formulas for the generalized zeta function and the generalized Ihara zeta function corresponding to the Grover walk and the positive-support version of the Grover walk on the regular graph via the Konno-Sato theorem, respectively. This paper extends these walks to a class of walks including random walks, correlated random walks, quantum walks, and open quantum random walks on the torus by the Fourier analysis. 
\end{abstract}

\vspace{10mm}

\begin{small}
\end{small}

\vspace{10mm}

\section{Introduction \label{sec01}}
In our previous paper \cite{KomatsuEtAl2021}, we studied a relation between the Grover walk and the zeta function based on the Konno-Sato theorem \cite{KonnoSato} and called this relation ``Grover/Zeta Correspondence". More precisely, we gave explicit formulas for the generalized zeta function and the generalized Ihara zeta function corresponding to the Grover walk with F-type and the positive-support version of the Grover walk with F-type on the vertex-transitive regular graph by the Konno-Sato theorem, respectively. The Grover walk is one of the most well-investigated quantum walks (QWs) inspired by the famous Grover algorithm. The QW is a quantum counterpart of the correlated random walk (CRW) which has the random walk (RW) as a special model. In fact, the CRW is the RW with memory. As for the QW, see \cite{Konno2008, ManouchehriWang, Portugal, Venegas} and as for the CRW and the RW, see \cite{Konno2009, Spitzer}, for example.

In this paper, we extend the Grover walk with F-type and the positive-support version of the Grover walk with F-type to a class of walks with both F- and M-types by using not the Konno-Sato theorem but a method of the Fourier transform for the case of the $d$-dimensional torus. Our class contains QWs and CRWs. Moreover, we can treat the open quantum random walk (OQRW) which has the CRW as a special model. Concerning the OQRW, see \cite{AttalEtAl2012b, AttalEtAl2012a}, for example. We call this kind of the zeta function the {\em walk-type} zeta function and call such a relationship ``Walk/Zeta Correspondence", corresponding to the above mentioned ``Grover/Zeta Correspondence". For the convenience of readers, we give a brief review of Grover/Zeta Correspondence presented by our previous paper \cite{KomatsuEtAl2021} in Appendix A.

The rest of this paper is organized as follows. In Section \ref{sec02}, we define the $2d$-state discrete time walk on the $d$-dimensional torus. Moreover, we explain a method of the Fourier transform for the walk. Section \ref{sec03} introduces the walk-type zeta function and presents our main results. In Section \ref{sec04}, we consider walks on the one-dimensional torus and give important models such as QWs, CRWs, RWs, and OQRWs. In Section \ref{secex}, we treat specific examples for QWs and CRWs. Section \ref{sec05} is devoted to the two-dimensional model. Finally, in Section \ref{sec06}, we deal with the higher-dimensional model. One of the interesting future problems might be to extend the torus to a suitable class of graphs for which we can use a similar method based on the Fourier analysis.

\section{Walk on Torus \label{sec02}}
First we introduce the following notation: $\mathbb{Z}$ is the set of integers, $\mathbb{Z}_{\ge}$ is the set of non-negative integers, $\mathbb{Z}_{>}$ is the set of positive integers, $\mathbb{R}$ is the set of real numbers, and $\mathbb{C}$ is the set of complex numbers. 

In this section, we give the definition of the $2d$-state discrete time walk on the {\em $d$-dimensional torus} $(d \geq 1)$ with $N^d$ vertices, denoted by $T^d_N$, where $N \in \mathbb{Z}_{>}$. Then we note that $T^d_N = (\mathbb{Z} \ \mbox{mod}\ N)^{d}$. The discrete time walk is defined by using a {\em shift operator} and a {\em coin matrix} which will be mentioned below.

Let $f : T^d_N \longrightarrow \mathbb{C}^{2d}$. For $j = 1,2,\ldots,d$ and $\xvec \in T^d_N$, the shift operator $\tau_j$ is defined by 
\begin{align*}
(\tau_j f)(\xvec) = f(\xvec-\evec_{j}),
\end{align*} 
where $\{ \evec_1,\evec_2,\ldots,\evec_d \}$ denotes the standard basis of $T^d_N$.

Let $A=[a_{ij}]_{i,j=1,2,\ldots,2d}$ be a $2d \times 2d$ matrix with $a_{ij} \in \mathbb{C}$ for $i,j =1,2,\ldots,2d$. We call $A$ the {\em coin matrix}. If $a_{ij} \in [0,1]$ and $\sum_{i=1}^{2d} a_{ij} = 1$ for any $j=1,2, \ldots, 2d$, then the walk is a CRW. In particular, when $a_{i1} = a_{i2} = \cdots = a_{i 2d}$ for any $i=1,2, \ldots, 2d$, this CRW becomes a RW. If $A$ is unitary, then the walk is a QW. So our class of walks contains RWs, CRWs, and QWs as special models.

To describe the evolution of the walk, we decompose the $2d \times 2d$ coin matrix $A$ as
\begin{align*}
A=\sum_{j=1}^{2d} P_{j} A,
\end{align*}
where $P_j$ denotes the orthogonal projection onto the one-dimensional subspace $\mathbb{C}\eta_j$ in $\mathbb{C}^{2d}$. Here $\{\eta_1,\eta_2, \ldots, \eta_{2d}\}$ denotes a standard basis on $\mathbb{C}^{2d}$.

The discrete time walk associated with the coin matrix $A$ on $T^d_N$ is determined by the $2d N^d \times 2d N^d$ matrix
\begin{align}
M_A=\sum_{j=1}^d \Big( P_{2j-1} A \tau_{j}^{-1} + P_{2j} A \tau_{j} \Big).
\label{unitaryop1}
\end{align}
The state at time $n \in \mathbb{Z}_{\ge}$ and location $\xvec \in T^d_N$ can be expressed by a $2d$-dimensional vector:
\begin{align*}
\Psi_{n}(\xvec)=
\begin{bmatrix}
\Psi^{1}_{n}(\xvec) \\ \Psi^{2}_{n}(\xvec) \\ \vdots \\ \Psi^{2d}_{n}(\xvec) 
\end{bmatrix} 
\in \mathbb{C}^{2d}.
\end{align*}

For $\Psi_n : T^d_N \longrightarrow \mathbb{C}^{2d} \ (n \in \mathbb{Z}_{\geq})$, from Eq. \eqref{unitaryop1}, the evolution of the walk is defined by 
\begin{align}
\Psi_{n+1}(\xvec) \equiv (M_A \Psi_{n})(\xvec)=\sum_{j=1}^{d}\Big(P_{2j-1}A\Psi_{n}(\xvec+\evec_j)+P_{2j}A\Psi_{n}(\xvec-\evec_j)\Big).
\label{reunitaryop1}
\end{align} 
This equation means that the walker moves at each step one unit to the $- x_j$-axis direction with matrix $P_{2j-1}A$ or one unit to the $x_j$-axis direction with matrix $P_{2j}A$ for $j=1,2, \ldots, d$. Moreover, for $n \in \ZM_{>}$ and $\xvec = (x_1, x_2, \ldots, x_d) \in T^d_N$, the $2d \times 2d$ matrix $\Phi_n (x_1, x_2, \ldots, x_d)$ is given by 
\begin{align}
\Phi_n (x_1, x_2, \ldots, x_d) = \sum_{\ast} \Xi_n \left(l_1,l_2, \ldots , l_{2d-1}, l_{2d} \right),
\label{xisum}
\end{align} 
where the $2d \times 2d$ matrix $\Xi_n \left(l_1,l_2, \ldots , l_{2d-1}, l_{2d} \right)$ is the sum of all possible paths in the trajectory of $l_{2j-1}$ steps $- x_j$-axis direction and  $l_{2j}$ steps $x_j$-axis direction and $\sum_{\ast}$ is the summation over $\left(l_1,l_2, \ldots , l_{2d-1}, l_{2d} \right) \in (\ZM_{\ge})^{2d}$ satisfying 
\begin{align*}
l_1 + l_2 + \cdots + l_{2d-1} + l_{2d} = n, \qquad x_j = - l_{2j-1} + l_{2j} \quad (j=1,2, \ldots, d).
\end{align*} 
For example, when $d=2$, $n=2$, and $(x_1,x_2)=(0,0)$, we see
\begin{align*}
\Phi_2 (0,0) = \Xi_2 \left(1,1,0,0 \right) + \Xi_2 \left(0,0,1,1 \right),
\end{align*} 
and
\begin{align*}
\Xi_2 \left(1,1,0,0 \right) 
&= (P_1 A)(P_2 A) + (P_2 A)(P_1 A), \quad
\\ 
\Xi_2 \left(0,0,1,1 \right) 
&= (P_3 A)(P_4 A) + (P_4 A)(P_3 A).
\end{align*} 
Here we put
\begin{align*}
\Phi_0 (x_1, x_2, \ldots, x_d)
= \left\{ 
\begin{array}{ll}
I_{2d} & \mbox{if $(x_1, x_2, \ldots, x_d) = (0, 0, \ldots, 0)$, } \\
O_{2d} & \mbox{if $(x_1, x_2, \ldots, x_d) \not= (0, 0, \ldots, 0)$},
\end{array}
\right.
\end{align*}
where $I_n$ is the $n \times n$ identity matrix and $O_n$ is the $n \times n$ zero matrix. Then, for the walk starting from $(0,0, \ldots, 0)$, we obtain  
\begin{align*}
\Psi_n (x_1, x_2, \ldots, x_d) = \Phi_n (x_1, x_2, \ldots, x_d) \Psi_0 (0, 0, \ldots, 0) \qquad (n \in \mathbb{Z}_{\ge}).
\end{align*} 
We call $\Phi_n (\xvec) = \Phi_n (x_1, x_2, \ldots, x_d)$ matrix weight at time $n \in \mathbb{Z}_{\ge}$ and location $\xvec \in T^d_N$ starting from ${\bf 0} = (0,0, \ldots, 0)$.

When we consider the walk on not $T^d_N$ but $\ZM^d$, we add the superscript ``$(\infty)$" to the notation like $\Psi^{(\infty)}$ and $\Xi^{(\infty)}$ which will be used in Sections \ref{sec03} and \ref{sec04}.

This type is {\em moving} shift model called {\em M-type} here. Another type is {\em flip-flop} shift model called {\em F-type} whose coin matrix is given by 
\begin{align*}
A^{(f)} = \left( I_d \otimes \sigma \right) A,
\end{align*} 
where $\otimes$ is the tensor product and
\begin{align*}
\sigma 
=
\begin{bmatrix}
0 & 1 \\ 
1 & 0 
\end{bmatrix} 
. 
\end{align*}
For example, when $d=3$, we have
\begin{align*}
I_3 \otimes \sigma = 
\begin{bmatrix}
\sigma & O_2 & O_2 \\ 
O_2 & \sigma & O_2 \\
O_2 & O_2 & \sigma
\end{bmatrix} 
.
\end{align*} 
The F-type model is also important, since it has a central role in the Konno-Sato theorem. When we distinguish $A$ (M-type) from $A^{(f)}$ (F-type), we write $A$ by $A^{(m)}$.

For time $n \in \mathbb{Z}_{\ge}$ and location $\xvec \in T^d_N$, we define the measure $\mu_n (\xvec)$ by
\begin{align*}
\mu_n (\xvec) = \| \Psi_n (\xvec) \|_{\mathbb{C}^{2d}}^p,
\end{align*} 
where $\|\cdot\|_{\mathbb{C}^{2d}}^p$ denotes the standard $p$-norm on $\mathbb{C}^{2d}$. That is,
\begin{align*}
\mu_n(\xvec) = \sum_{j=1}^{2d}|\Psi_n^{j}(\xvec)|^p.
\end{align*}
As for CRWs and QWs, we take $p=1$ and $p=2$, respectively. Then CRWs and QWs satisfy 
\begin{align*}
\sum_{\xvec \in T_N^d} \mu_n (\xvec) = \sum_{\xvec \in T_N^d} \mu_0 (\xvec), 
\end{align*} 
for any time $n \in \ZM_{>}$. However, we do not necessarily impose such a condition for the walk we consider here. For example, the two-dimensional positive-support version of the Grover walk (introduced in Section \ref{sec05}) does not satisfy the condition. In this meaning, our walk is a generalized version for the usual walk.

To consider the zeta function, we use the Fourier analysis. To do so, we introduce the following notation: $\mathbb{K}_N = \{ 0,1, \ldots, N-1 \}$ and $\widetilde{\mathbb{K}}_N = \{ 0 ,2 \pi/N, \ldots, 2 \pi (N-1)/N \}$.

For $f : \mathbb{K}_N^d \longrightarrow \mathbb{C}^{2d}$, the Fourier transform of the function $f$, denoted by $\widehat{f}$, is defined by the sum
\begin{align}
\widehat{f}(\kvec) = \frac{1}{N^{d/2}} \sum_{\xvec \in \mathbb{K}_N^d} e^{- 2 \pi i \langle \xvec, \kvec \rangle /N} \ f(\xvec),
\label{yoiko01}
\end{align}
where $\kvec=(k_1,k_2,\ldots,k_{d}) \in \mathbb{K}_N^d$. Here $\langle \xvec,  \kvec \rangle$ is the canonical inner product of $\mathbb{R}^d$, i.e., $\langle \xvec,  \kvec \rangle = \sum_{j=1}^{d} x_j k_j$. Then we see that $\widehat{f} : \mathbb{K}_N^d \longrightarrow \mathbb{C}^{2d}$. Moreover, we should remark that 
\begin{align}
f(\xvec) = \frac{1}{N^{d/2}} \sum_{\kvec \in \mathbb{K}_N^d} e^{2 \pi i \langle \xvec, \kvec \rangle /N} \ \widehat{f}(\kvec),
\label{yoiko02}
\end{align}
where $\xvec =(x_1,x_2,\ldots,x_{d}) \in \mathbb{K}_N^d$. By using 
\begin{align}
\widetilde{k}_j = \frac{2 \pi k_j}{N} \in \widetilde{\mathbb{K}}_N, \quad \widetilde{\kvec}=(\widetilde{k}_1,\widetilde{k}_2,\ldots,\widetilde{k}_{d}) \in \widetilde{\mathbb{K}}_N^d, 
\label{kantan01}
\end{align}
we can rewrite Eqs. \eqref{yoiko01} and \eqref{yoiko02} in the following way: 
\begin{align}
\widehat{g}(\widetilde{\kvec}) 
&= \frac{1}{N^{d/2}} \sum_{\xvec \in \mathbb{K}_N^d} e^{- i \langle \xvec, \widetilde{\kvec} \rangle} \ g(\xvec),
\nonumber
\\
g(\xvec) 
&= \frac{1}{N^{d/2}} \sum_{\widetilde{\kvec} \in \widetilde{\mathbb{K}}_N^d} e^{i \langle \xvec, \widetilde{\kvec} \rangle} \ \widehat{g}(\widetilde{\kvec}),
\label{burahu02}
\end{align}
for $g : \mathbb{K}_N^d \longrightarrow \mathbb{C}^{2d}$ and $\widehat{g} : \widetilde{\mathbb{K}}_N^d \longrightarrow \mathbb{C}^{2d}$. In order to take a limit $N \to \infty$, we introduced the notation given in Eq. \eqref{kantan01}. We should note that as for the summation, we sometimes write ``$\kvec \in \mathbb{K}_N^d$" instead of ``$\widetilde{\kvec} \in \widetilde{\mathbb{K}}_N^d$", for example, 
\begin{align*}
g(\xvec) 
= \frac{1}{N^{d/2}} \sum_{\kvec \in \mathbb{K}_N^d} e^{i \langle \xvec, \widetilde{\kvec} \rangle} \ \widehat{g}(\widetilde{\kvec}),
\end{align*}
instead of Eq. \eqref{burahu02}.

From the Fourier transform and Eq. \eqref{reunitaryop1}, we have
\begin{align*}
\widehat{\Psi}_{n+1}(\kvec)=\widehat{M}_A(\kvec)\widehat{\Psi}_n(\kvec),
\end{align*}
where $\Psi_n : T_N^d \longrightarrow \mathbb{C}^{2d}$ and $2d \times 2d$ matrix $\widehat{M}_A(\kvec)$ is determined by
\begin{align*}                          
\widehat{M}_A(\kvec)=\sum_{j=1}^{d} \Big( e^{2 \pi i k_j/N} P_{2j-1} A + e^{-2 \pi i k_j /N} P_{2j} A \Big). 
\end{align*}
By using notations in Eq. \eqref{kantan01}, we have 
\begin{align}                          
\widehat{M}_A(\widetilde{\kvec})=\sum_{j=1}^{d} \Big( e^{i \widetilde{k}_j} P_{2j-1} A + e^{-i \widetilde{k}_j} P_{2j} A \Big). 
\label{migiyoshi01}
\end{align}
Next we will consider the following eigenvalue problem for $2d N^d \times 2d N^d$ matrix $M_A$: 
\begin{align}
\lambda \Psi = M_A \Psi, 
\label{yoiko03}
\end{align}
where $\lambda \in \mathbb{C}$ is an eigenvalue and $\Psi (\in \mathbb{C}^{2d N^d})$ is the corresponding eigenvector. Noting that Eq. \eqref{yoiko03} is closely related to Eq. \eqref{reunitaryop1}, we see that Eq. \eqref{yoiko03} is rewritten as 
\begin{align}
\lambda \Psi (\xvec) = (M_A \Psi)(\xvec) = \sum_{j=1}^{d}\Big(P_{2j-1} A \Psi (\xvec+\evec_j)+P_{2j} A \Psi (\xvec-\evec_j)\Big),
\label{yoiko04}
\end{align} 
for any $\xvec \in \mathbb{K}_N^d$. From the Fourier transform and Eq. \eqref{yoiko04}, we have
\begin{align*}
\lambda \widehat{\Psi} (\kvec) = \widehat{M}_A (\kvec) \widehat{\Psi} (\kvec), 
\end{align*}
for any $\kvec \in \mathbb{K}_N^d$. Then the characteristic polynomials of $2d \times 2d$ matrix $\widehat{M}_A(\kvec)$ for fixed $\kvec (\in \mathbb{K}_N^d)$ is 
\begin{align}                          
\det \Big( \lambda I_{2d} - \widehat{M}_A (\kvec) \Big) = \prod_{j=1}^{2d} \Big( \lambda - \lambda_{j} (\kvec) \Big),
\label{yoiko06}
\end{align}
where $\lambda_{j} (\kvec)$ are eigenvalues of $\widehat{M}_A (\kvec)$. Similarly, the characteristic polynomials of $2d N^d \times 2d N^d$ matrix $\widehat{M}_A$ is 
\begin{align*}                          
\det \Big( \lambda I_{2d N^d} - \widehat{M}_A  \Big) = \prod_{j=1}^{2d} \prod_{\kvec \in \mathbb{K}_N^d} \Big( \lambda - \lambda_{j} (\kvec) \Big).
\end{align*}
Thus we have
\begin{align*}                          
\det \Big( \lambda I_{2d N^d} - M_A  \Big) 
&= \det \Big( \lambda I_{2d N^d} - \widehat{M}_A  \Big) = \prod_{j=1}^{2d} \prod_{\kvec \in \mathbb{K}_N^d} \Big( \lambda - \lambda_{j} (\kvec) \Big).
\end{align*}
Therefore, by taking $\lambda = 1/u$, we have the following key result.
\begin{lemma}  
\begin{align*}                          
\det \Big( I_{2d N^d} - u M_A  \Big) 
= \det \Big( I_{2d N^d} - u \widehat{M}_A  \Big) = \prod_{j=1}^{2d} \prod_{\kvec \in \mathbb{K}_N^d} \Big( 1 - u \lambda_{j} (\kvec) \Big).
\end{align*}
\label{keyresult01lemma}
\end{lemma}  
We should note that for fixed $\kvec (\in \mathbb{K}_N^d)$, eigenvalues of  $2d \times 2d$ matrix $\widehat{M}_A (\kvec)$ are expressed as 
\begin{align*}
{\rm Spec} ( \widehat{M}_A (\kvec)) = \left\{  \lambda_{j} (\kvec) \ | \ j = 1, 2, \ldots, 2d \right\}. 
\end{align*}
Moreover, eigenvalues of $2d N^d \times 2d N^d$ matrix not only $\widehat{M}_A$ but also $M_A$ are expressed as 
\begin{align*}
{\rm Spec} ( \widehat{M}_A) = {\rm Spec} (M_A)= \left\{  \lambda_{j} (\kvec) \ | \ j = 1, 2, \ldots, 2d, \ \kvec \in \mathbb{K}_N^d \right\}. 
\end{align*}


By using notations in Eq. \eqref{kantan01} and Eq. \eqref{yoiko06}, we see that for fixed $\kvec (\in \mathbb{K}_N^d)$,  
\begin{align}                          
\det \Big( I_{2d} - u \widehat{M}_A (\widetilde{\kvec}) \Big) = \prod_{j=1}^{2d} \Big( 1 - u \lambda_{j} (\widetilde{\kvec}) \Big).
\label{yoiko08}
\end{align}
Moreover, from Eq. \eqref{migiyoshi01}, we have the following important formula.
\begin{lemma}
\begin{align*}                          
\det \Big( I_{2d} - u \widehat{M}_A (\widetilde{\kvec}) \Big) = \det \left(I_{2d} - u \times \sum_{j=1}^{d} \Big( e^{i \widetilde{k}_j} P_{2j-1} A + e^{-i \widetilde{k}_j} P_{2j} A \Big) \right). 
\end{align*}
\label{lemmakoreda}
\end{lemma}

\section{Walk-Type Zeta Function \label{sec03}}
For our setting in the previous section, we define the {\em walk-type zeta function} by 
\begin{align}
\overline{\zeta} \left(A, T^d_N, u \right) = \det \Big( I_{2d N^d} - u M_A \Big)^{-1/N^d}.
\label{satosan01}
\end{align}
In general, for a $d_c \times d_c$ coin matrix $A$, we put  
\begin{align*}
\overline{\zeta} \left(A, T^d_N, u \right) = \det \Big(I_{d_c N^d} - u M_A \Big)^{-1/N^d}.
\end{align*}
We should remark that the walk-type zeta function becomes the generalized zeta function $\overline{\zeta} \left(T^d_N, u \right)$ in \cite{KomatsuEtAl2021} for the Grover walk (F-type). See also Appendix A. So we write the walk-type zeta function with a coin matrix $A$ as $\overline{\zeta} \left(A, T^d_N, u \right)$. Furthermore, our walk is defined on the ``site" $\xvec (\in T^d_N)$, on the other hand, the walk in \cite{KomatsuEtAl2021} is defined on the ``arc" (i.e., directed edge). However both of the walks are same for the torus case.

We here briefly explain zeta functions related to our walk. As for the zeta function, see \cite{KomatsuEtAl2021, KonnoSato}, for example. Starting from $p$-adic Selberg zeta functions, Ihara \cite{Ihara} introduced the Ihara zeta functions of graphs, and showed that the reciprocals of the Ihara zeta functions of regular graphs are explicit polynomials. Recently, the Ihara zeta function of a finite graph was extended to an infinite graph. Clair \cite{Clair} computed the Ihara zeta function for the infinite grid by using elliptic integrals and theta functions. Chinta et al. \cite{ChintaEtAl} obtained a generalized version of the determinant formula for the Ihara zeta function associated to finite or infinite graphs. 

By Lemma \ref{keyresult01lemma} and Eqs. \eqref{yoiko08} and \eqref{satosan01}, we compute
\begin{align*}
\overline{\zeta} \left(A, T^d_N, u \right) ^{-1} 
&= \exp \left[ \frac{1}{N^d} \log \left\{ \det \Big( I_{2d N^d} - u M_A \Big) \right\} \right]
\\
&= \exp \left[ \frac{1}{N^d} \log \left\{ \det \Big( I_{2d N^d} - u \widehat{M}_A \Big) \right\} \right]
\\
&= \exp \left[ \frac{1}{N^d} \sum_{\widetilde{\kvec} \in \widetilde{\mathbb{K}}_N^d} \log \left\{ \det \Big( I_{2d} - u \widehat{M}_A (\widetilde{\kvec}) \Big) \right\} \right].
\end{align*}
So we have 
\begin{align*}
\overline{\zeta} \left(A, T^d_N, u \right) ^{-1}
=
\exp \left[ \frac{1}{N^d} \sum_{\widetilde{\kvec} \in \widetilde{\mathbb{K}}_N^d} \log \left\{ \det \Big( I_{2d} - u \widehat{M}_A (\widetilde{\kvec}) \Big) \right\} \right].
\end{align*}
Sometimes we write $\sum_{\kvec \in \mathbb{K}_N^d}$ instead of $\sum_{\widetilde{\kvec} \in \widetilde{\mathbb{K}}_N^d}$ as follows:
\begin{align*}
\overline{\zeta} \left(A, T^d_N, u \right) ^{-1}
=
\exp \left[ \frac{1}{N^d} \sum_{\kvec \in \mathbb{K}_N^d} \log \left\{ \det \Big( I_{2d} - u \widehat{M}_A (\widetilde{\kvec}) \Big) \right\} \right].
\end{align*}
Noting $\widetilde{k_j} = 2 \pi k_j/N \ (j=1,2, \ldots, d)$ and taking a limit as $N \to \infty$, we obtain
\begin{align*}
\lim_{N \to \infty} \overline{\zeta} \left(A, T^d_N, u \right) ^{-1}
=
\exp \left[ \int_{[0,2 \pi)^d} \log \left\{ \det \Big( I_{2d} - u \widehat{M}_A \left( \Theta^{(d)} \right) \Big) \right\} d \Theta^{(d)}_{unif} \right],
\end{align*}
if the limit exists. We should note that when we take a limit as $N \to \infty$, we assume that the limit exists throughout this paper. Here $\Theta^{(d)} = (\theta_1, \theta_2, \ldots, \theta_d) (\in [0, 2 \pi)^d)$ and $d \Theta^{(d)}_{unif}$ denotes the uniform measure on $[0, 2 \pi)^d$, that is,
\begin{align*}
d \Theta^{(d)}_{unif} = \frac{d \theta_1}{2 \pi } \cdots \frac{d \theta_d}{2 \pi }.
\end{align*}
Therefore we obtain one of our main results.
\begin{theorem}
\begin{align*}
\overline{\zeta} \left(A, T^d_N, u \right) ^{-1}
&= \exp \left[ \frac{1}{N^d} \sum_{\widetilde{\kvec} \in \widetilde{\mathbb{K}}_N^d} \log \left\{ \det \Big( F(\widetilde{\kvec}, u) \Big) \right\} \right],
\\
\lim_{N \to \infty} \overline{\zeta} \left(A, T^d_N, u \right) ^{-1}
&=
\exp \left[ \int_{[0,2 \pi)^d} \log \left\{ \det \Big( F \left( \Theta^{(d)}, u \right)  \Big) \right\} d \Theta^{(d)}_{unif} \right],
\end{align*}
where 
\begin{align*}
F \left( \wvec , u \right) = I_{2d} - u \widehat{M}_A (\wvec), 
\end{align*}
with $\wvec = (w_1, w_2, \ldots, w_d) \in \RM^d$.
\label{th001}
\end{theorem}

Furthermore, we define $C_r (A, T^d_N)$ by
\begin{align}
\overline{\zeta} \left(A, T^d_N, u \right) = \exp \left( \sum_{r=1}^{\infty} \frac{C_r (A, T^d_N)}{r} u^r \right).
\label{satosan03}
\end{align}
Sometime we write $C_r (A, T^d_N)$ by $C_r$ for short. Combining Eq. \eqref{satosan01} with Eq. \eqref{satosan03} implies 
\begin{align*}
\det \Big( I_{2d N^d} - u M_A \Big)^{-1/N^d} = \exp \left( \sum_{r=1}^{\infty} \frac{C_r}{r} u^r \right).
\end{align*}
Thus we get
\begin{align}
- \frac{1}{N^d} \log \left\{ \det \Big( I_{2d N^d} - u M_A \Big) \right\} =  \sum_{r=1}^{\infty} \frac{C_r}{r} u^r.
\label{satosan05}
\end{align}
It follows from Lemma \ref{keyresult01lemma} that the left-hand of Eq. \eqref{satosan05} becomes 
\begin{align*}
- \frac{1}{N^d} \log \left\{ \det \Big( I_{2d N^d} - u M_A \Big) \right\} &=
- \frac{1}{N^d} \log \left\{ \det \Big( I_{2d N^d} - u \widehat{M}_A \Big) \right\} 
\\
&= 
- \frac{1}{N^d} \sum_{j=1}^{2d} \sum_{\kvec \in \mathbb{K}_N^d} \log \left\{ 1 - u \lambda_{j} (\kvec) \right\} 
\\
&= \frac{1}{N^d} \sum_{j=1}^{2d} \sum_{\kvec \in \mathbb{K}_N^d} \sum_{r=1}^{\infty} \frac{\left(\lambda_{j} (\kvec) \right)^r}{r} u^r. 
\end{align*}
By this and the right-hand of Eq. \eqref{satosan05}, we have
\begin{align*}
C_r (A, T^d_N) = \frac{1}{N^d} \sum_{j=1}^{2d} \sum_{\kvec \in \mathbb{K}_N^d} \left(\lambda_{j} (\kvec) \right)^r
\end{align*}
This is rewritten as 
\begin{align}
C_r (A, T^d_N) = \frac{1}{N^d} \sum_{j=1}^{2d} \sum_{\widetilde{\kvec} \in \widetilde{\mathbb{K}}_N^d} \left(\lambda_{j} (\widetilde{\kvec}) \right)^r.
\label{satosan06t}
\end{align}
Noting $\widetilde{k_j} = 2 \pi k_j/N \ (j=1,2, \ldots, d)$ and taking a limit as $N \to \infty$, we get
\begin{align}
\lim_{N \to \infty} C_r (A, T^d_N) = \sum_{j=1}^{2d} \int_{[0,2 \pi)^d} \lambda_{j} \left( \Theta^{(d)} \right)^r d \Theta^{(d)}_{unif}.
\label{satosan06limit}
\end{align}
Let ${\rm Tr} (A)$ denote the trace of a square matrix $A$. Therefore by definition of ${\rm Tr}$ and Eqs. \eqref{satosan06t} and \eqref{satosan06limit}, we obtain
\begin{prop}
\begin{align}
C_r (A, T^d_N) 
&
= \frac{1}{N^d} \sum_{\widetilde{\kvec} \in \widetilde{\mathbb{K}}_N^d} {\rm Tr} \left( \left( \widehat{M}_A (\widetilde{\kvec}) \right)^r \right),
\nonumber
\\
\lim_{N \to \infty} C_r (A, T^d_N) 
&
= \int_{[0,2 \pi)^d} {\rm Tr} \left( \left( \widehat{M}_A (\Theta^{(d)}) \right)^r \right) d \Theta^{(d)}_{unif}.
\label{satosan07}
\end{align}
\label{satosan06prop}
\end{prop}

Furthermore, noting that
\begin{align*}
\int_{[0,2 \pi)^d} e^{ i \left( m_1 \theta_1 + m_2 \theta_2 + \cdots + m_d \theta_d \right)}  d \Theta^{(d)}_{unif} = 0, 
\end{align*}
for $(m_1, m_2, \ldots, m_d) \in \ZM^d$ with $(m_1, m_2, \ldots, m_d) \not= (0,0, \ldots, 0)$, we have
\begin{align}
\int_{[0,2 \pi)^d} \left( \widehat{M}_A (\Theta^{(d)}) \right)^r d \Theta^{(d)}_{unif} = \sum_{\ast} \Xi_r ^{(\infty)} \left( l_1, l_2, \ldots, l_{2d-1}, l_{2d} \right),
\label{satosan09}
\end{align}
where $\sum_{\ast}$ is the summation over $\left(l_1,l_2, \ldots , l_{2d-1}, l_{2d} \right)$ satisfying 
\begin{align*}
l_1 + l_2 + \cdots + l_{2d-1} + l_{2d} = r, \qquad l_{2j-1} = l_{2j} \quad (j=1,2, \ldots, 2d).
\end{align*} 
From Eqs. \eqref{xisum}, \eqref{satosan07}, and \eqref{satosan09}, we obtain one of our main results.
\begin{theorem}
\begin{align*}
\lim_{N \to \infty} C_r (A, T^d_N) = {\rm Tr} \left( \Phi_r ^{(\infty)} ({\bf 0}) \right),
\end{align*}
where ${\bf 0} = (0,0, \ldots, 0)$.
\label{satosan10thm}
\end{theorem}
An interesting point is that $\Phi_r ^{(\infty)} ({\bf 0})$ is the return ``matrix weight" at time $r$ for the walk on not $T_N^d$ but $\ZM^d$. We should remark that in general ${\rm Tr} ( \Phi_r ^{(\infty)} ({\bf 0}) )$ is not the same as the return probability at time $r$ for the walk (which will be briefly explained in Sections \ref{sec04} and \ref{secex}).

To understand the derivation of Theorem \ref{satosan10thm} well, we consider the one-dimensional model with a coin matrix $A$ as
\begin{align*}
A =
\begin{bmatrix}
a_{11} & a_{12} \\ 
a_{21} & a_{22} 
\end{bmatrix} 
.
\end{align*}
In a similar way, we can extend this argument to general $d$-dimensional model. We begin with $r=1, 2$ cases for $\left( \widehat{M}_{A} (\theta) \right)^r$. Then we have
\begin{align*} 
\widehat{M}_{A} (\theta)
&= e^{i \theta} P_{1} A + e^{-i \theta} P_{2} A,
\\
\left( \widehat{M}_{A} (\theta) \right)^2
&= e^{2 i \theta} (P_{1} A)^2 + \left\{ (P_{1} A) (P_{2} A) + (P_{2} A) (P_{1} A) \right\} + e^{-2i \theta} (P_{2} A)^2.
\end{align*}
Thus we see
\begin{align*} 
\int_0^{2 \pi} \widehat{M}_{A} (\theta) \frac{d \theta}{2 \pi}
&= \int_0^{2 \pi} \left(  e^{i \theta} P_{1} A + e^{-i \theta} P_{2} A \right) \frac{d \theta}{2 \pi}
= O_2,
\\
\int_0^{2 \pi} \left( \widehat{M}_{A} (\theta) \right)^2 \frac{d \theta}{2 \pi} 
&= \int_0^{2 \pi} \left( e^{2 i \theta} (P_{1} A)^2 + \left\{ (P_{1} A) (P_{2} A) + (P_{2} A) (P_{1} A) \right\} + e^{-2i \theta} (P_{2} A)^2 \right) \frac{d \theta}{2 \pi}
\\
&= (P_{1} A) (P_{2} A) + (P_{2} A) (P_{1} A) 
\\
&= \Xi_2^{(\infty)} (1,1) = \Phi_2^{(\infty)} (0).
\end{align*}
A similar argument implies that for $l=1,2, \ldots$, we have
\begin{align*}
\int_0^{2 \pi} \left( \widehat{M}_{A} (\theta) \right)^r \frac{d \theta}{2 \pi} = \left\{
\begin{array}{ll}
O_2  & \mbox{if $r=2l-1$, } \\
\Xi_{2l}^ {(\infty)} (l,l) & \mbox{if $r=2l$}.
\end{array}
\right.
\end{align*}
By definition of $\Phi_{r}^{(\infty)} (0)$, we see 
\begin{align*}
\Phi_{r}^{(\infty)} (0) 
= \left\{ 
\begin{array}{ll}
O_2 & \mbox{if $r=2l-1$, } \\
\Xi_{2l}^{(\infty)} (l,l) & \mbox{if $r=2l$}.
\end{array}
\right.
\end{align*}
So we immediately get
\begin{align*}
\int_0^{2 \pi} \left( \widehat{M}_{A} (\theta) \right)^r \frac{d \theta}{2 \pi} = \Phi_{r}^{(\infty)} (0).
\end{align*}
Thus we have the desired conclusion for the one-dimensional case:
\begin{align*}
\lim_{N \to \infty} C_r (A, T^1_N) 
= \int_0^{2 \pi} {\rm Tr} \left( \left( \widehat{M}_A (\theta) \right)^r \right) \frac{d \theta}{2 \pi} =  {\rm Tr} \left( \Phi_{r} ^{(\infty)} (0) \right).
\end{align*}
On the other hand, we obtained the following expression $\Phi_{r} ^{(\infty)} (0)$ for $r=2l$ (see Konno \cite{Konno2008, Konno2009}, for example).
\begin{lemma}
We consider the walk with a coin matrix $A$ on $\ZM$. Assume that $a_{11} a_{12} a_{21} a_{22} \not=0$. Then for $l=1,2, \ldots$, we have 
\begin{align*}
\Phi_{2l} ^{(\infty)} (0) 
&= \Xi_{2l} ^{(\infty)} \left( l,l \right) 
\\
&= \left( a_{11} a_{22} \right)^l \sum_{m=1}^l \left( \frac{a_{12} a_{21}}{a_{11} a_{22}} \right)^m {l-1 \choose m-1}^2 
\left[ \frac{l-m}{a_{11}m} Q_1 + \frac{l-m}{a_{22}m} Q_2 + \frac{1}{a_{12}} Q_3 +  \frac{1}{a_{21}} Q_4 \right],
\end{align*}
where
\begin{align*}
Q_1 = P_1 A, \quad Q_2 = P_2 A, \quad Q_3 = \sigma P_1 A, \quad Q_4 = \sigma P_2 A, 
\end{align*}
and 
\begin{align*}
A
=
\begin{bmatrix}
a_{11} & a_{12} \\ 
a_{21} & a_{22} 
\end{bmatrix} 
, \quad 
P_{1} 
=
\begin{bmatrix}
1 & 0 \\ 
0 & 0 
\end{bmatrix}
, \quad 
P_{2} 
=
\begin{bmatrix}
0 & 0 \\ 
0 & 1 
\end{bmatrix}
, \quad
\sigma
= 
\begin{bmatrix}
0 & 1 \\ 
1 & 0 
\end{bmatrix}
. 
\end{align*}
\label{tukaerupre}
\end{lemma}
We note that 
\begin{align*}
Q_1
=
\begin{bmatrix}
a_{11} & a_{12} \\ 
0 & 0 
\end{bmatrix} 
, \qquad 
Q_2
=
\begin{bmatrix}
0 & 0 \\
a_{21} & a_{22} 
\end{bmatrix} 
, \qquad
Q_3
=
\begin{bmatrix}
0 & 0 \\
a_{11} & a_{12} 
\end{bmatrix} 
, \qquad 
Q_4
=
\begin{bmatrix}
a_{21} & a_{22} \\ 
0 & 0
\end{bmatrix} 
.
\end{align*}
Moreover, a direct computation gives
\begin{align}
{\rm Tr} \left( \frac{l-m}{a_{11}m} Q_1 + \frac{l-m}{a_{22}m} Q_2 + \frac{1}{a_{12}} Q_3 +  \frac{1}{a_{21}} Q_4 \right) = \frac{2 l}{m}.
\label{matsuyama}
\end{align}
So finally Lemma \ref{tukaerupre} implies
\begin{lemma}
We consider the walk with a coin matrix $A$ on $T_N^1$. Assume that $a_{11} a_{12} a_{21} a_{22} \not=0$. Then for $l=1,2, \ldots$, we have 
\begin{align*}
\lim_{N \to \infty} C_{2l} (A, T^1_N) 
&= {\rm Tr} \left( \Phi_{2l} ^{(\infty)} (0) \right) = {\rm Tr} \left( \Xi_{2l} ^{(\infty)} \left( l,l \right) \right) 
\\
&= 2 l \left( a_{11} a_{22} \right)^l \sum_{m=1}^l \frac{1}{m} {l-1 \choose m-1}^2 \left( \frac{a_{12} a_{21}}{a_{11} a_{22}} \right)^m 
\\
&= 
2 l  \left( a_{11} a_{22} \right)^{l-1} \left(a_{12} a_{21} \right) {}_2F_1 \left( 1-l , 1-l ; 2 ; \frac{a_{12} a_{21}}{a_{11} a_{22}} \right), 
\\
\lim_{N \to \infty} C_{2l-1} (A, T^1_N) 
&= 0.
\end{align*}
\label{tukaeru}
\end{lemma}
Here we introduced the hypergeometric series ${}_2F_1(a, b; c ;z)$ (see Andrews et al. \cite{Andrews1999}, for example). In particular, we used the following relation:
\begin{align*}
\sum_{k=1}^n \frac{1}{k} {n-1 \choose k-1}^2 z^{k-1} = {}_2F_1 \left( 1-n , 1-n ; 2 ; z \right).
\end{align*}
If we take $z=1$, then we have 
\begin{align*}
\sum_{k=1}^n \frac{1}{k} {n-1 \choose k-1}^2  = {}_2F_1 \left( 1-n , 1-n ; 2 ; 1 \right) = \frac{\Gamma (2)\Gamma (2n)}{\Gamma (n+1)^2} = \frac{(2n-1)!}{(n!)^2}.
\end{align*}
where $\Gamma (z)$ is the gamma function. Thus we get the following result will be used in Corollary \ref{kimarid1crwrw} for the RW case.
\begin{align}
2n \sum_{k=1}^n \frac{1}{k} {n-1 \choose k-1}^2 ={2n \choose n}.
\label{atode}
\end{align}

In the next section, we will give explicit expressions of $\lim_{N \to \infty} C_r (A, T^1_N)$ for RWs, CRWs, and QWs by using Lemma \ref{tukaeru}.

\section{One-Dimensional Case \label{sec04}} 
This section deals with walks on the one-dimensional torus $T^1_N$ whose $2 \times 2$ coin matrix $A^{(m)}$ (M-type) or $A^{(f)}$ (F-type) as follows: 
\begin{align*}
A^{(m)}  
=
\begin{bmatrix}
a_{11} & a_{12} \\ 
a_{21} & a_{22} 
\end{bmatrix} 
, \qquad 
A^{(f)} 
=  
\begin{bmatrix}
a_{21} & a_{22} \\ 
a_{11} & a_{12} 
\end{bmatrix}
,
\end{align*}
since
\begin{align*}
A^{(f)} = \left( I_1 \otimes \sigma \right) A^{(m)} = \sigma  A^{(m)} = 
\begin{bmatrix}
0 & 1 \\ 
1 & 0 
\end{bmatrix}
\begin{bmatrix}
a_{11} & a_{12} \\ 
a_{21} & a_{22} 
\end{bmatrix} 
.
\end{align*}
The one-dimensional walk is the first example, so we will explain it in detail. Put $k=k_1$ and $\widetilde{k} = \widetilde{k}_1$. In this case, we take 
\begin{align*}
P_{1} 
=
\begin{bmatrix}
1 & 0 \\ 
0 & 0 
\end{bmatrix}
, \qquad 
P_{2} 
=
\begin{bmatrix}
0 & 0 \\ 
0 & 1 
\end{bmatrix}
. 
\end{align*}
Thus we immediately get
\begin{align} 
\widehat{M}_{A^{(m)}} (\widetilde{k})
&= e^{i \widetilde{k}} P_{1} A^{(m)} + e^{-i \widetilde{k}} P_{2} A^{(m)} 
= 
\begin{bmatrix}
e^{i \widetilde{k}} a_{11} & e^{i \widetilde{k}} a_{12} \\ 
e^{-i \widetilde{k}}a_{21} & e^{-i \widetilde{k}} a_{22} 
\end{bmatrix} 
,
\label{jinji01}
\\
\widehat{M}_{A^{(f)}} (\widetilde{k})
&= e^{i \widetilde{k}} P_{1} A^{(f)} + e^{-i \widetilde{k}} P_{2} A^{(f)} 
= 
\begin{bmatrix}
e^{i \widetilde{k}} a_{21} & e^{i \widetilde{k}} a_{22} \\ 
e^{-i \widetilde{k}}a_{11} & e^{-i \widetilde{k}} a_{12} 
\end{bmatrix} 
.
\label{jinji02}
\end{align}
By these equations, we have
\begin{align*}
\det \Big( I_{2} - u \widehat{M}_{A^{(s)}} (\widetilde{k}) \Big) 
= 1 - {\rm Tr} \left(  \widehat{M}_{A^{(s)}} (\widetilde{k}) \right) u + \det \left(  \widehat{M}_{A^{(s)}} (\widetilde{k}) \right) u^2 \qquad (s \in \{m,f\}).
\end{align*}
From Theorem \ref{th001}, we obtain
\begin{prop}
\begin{align*}
\overline{\zeta} \left(A^{(s)}, T^1_N, u \right)^{-1} 
&= \exp \left[ \frac{1}{N} \sum^{N-1}_{k =0} \log \left\{ 1 - {\rm Tr} \left(  \widehat{M}_{A^{(s)}}  (\widetilde{k}) \right) u + \det \left(  \widehat{M}_{A^{(s)}} (\widetilde{k}) \right) u^2 \right\} \right],
\\
\lim_{N \to \infty} \overline{\zeta} \left(A^{(s)}, T^1_N, u \right)^{-1} 
&= \exp \left[ \int_0^{2 \pi} \log \left\{ 1 - {\rm Tr} \left(  \widehat{M}_{A^{(s)}}  (\theta) \right) u + \det \left(  \widehat{M}_{A^{(s)}} (\theta) \right) u^2 \right\} \frac{d \theta}{2 \pi} \right],
\end{align*}
for $s \in \{m,f\}$.
\label{kimarid1}
\end{prop}
From now on, we consider QWs, CRWs, and RWs and apply Proposition \ref{kimarid1} and Lemma \ref{tukaeru} to their models.
\par
\
\par
(a) QW case. 
\par
\
\par
One of the typical classes for $2 \times 2$ coin matrix $A^{(m)}$ (M-type) or $A^{(f)}$ (F-type) is as follows: 
\begin{align*}
A^{(m)} 
=
\begin{bmatrix}
\cos \xi & \sin \xi  \\ 
\sin \xi & - \cos \xi  
\end{bmatrix} 
, \qquad 
A^{(f)} 
=  
\begin{bmatrix}
\sin \xi & - \cos \xi \\ 
\cos \xi & \sin \xi  
\end{bmatrix}
\qquad ( \xi \in [0, 2\pi)).
\end{align*}
When $\xi = \pi/4$, the QW becomes the so-called {\em Hadamard walk} which is one of the most well-investigated model in the study of QWs like the Grover walk. 

From Proposition \ref{kimarid1}, we have
\begin{cor} 
\begin{align*}
\overline{\zeta} \left(A^{(s)}, T^1_N, u \right)^{-1} 
&= \exp \left[ \frac{1}{N} \sum^{N-1}_{k =0} \log \left( F^{(s)} \left( \widetilde{k}, u \right)  \right) \right],
\\
\lim_{N \to \infty} \overline{\zeta} \left(A^{(s)}, T^1_N, u \right)^{-1} 
&= \exp \left[ \int_0^{2 \pi} \log \left( F^{(s)} \left( \theta, u \right) \right) \frac{d \theta}{2 \pi} \right],
\end{align*}
for $s \in \{m,f\}$, where
\begin{align*}
F^{(m)} \left( w, u \right) 
&= 1 - 2 i \cos \xi \sin w \cdot u - u^2,
\\
F^{(f)} \left( w, u \right) 
&= 1 - 2 \sin \xi \cos w \cdot u + u^2.
\end{align*}
\label{kimarid1qw}
\end{cor}
Furthermore, Lemma \ref{tukaeru} implies
\begin{cor} 
\begin{align*}
\lim_{N \to \infty} C_{2l} (A^{(m)}, T^1_N) 
&= 2 l \left(-  \cos^2 \xi \right)^{l} \sum_{m=1}^l \frac{1}{m} {l-1 \choose m-1}^2 \left( - \tan^2 \xi \right)^{m}
\\
&= 2 l \left(-  \cos^2 \xi \right)^{l-1} (\sin^2 \xi) \ {}_2F_1 \left( 1-l , 1-l ; 2 ; - \tan^2 \xi \right),
\\
\lim_{N \to \infty} C_{2l} (A^{(f)}, T^1_N) 
&= 2 l \left(\sin \xi \right)^{2l} \sum_{m=1}^l \frac{1}{m} {l-1 \choose m-1}^2 \left( - \cot^2 \xi \right)^{m}
\\
&= 2 l \left(\sin \xi \right)^{2(l-1)} (- \cos^2 \xi) \ {}_2F_1 \left( 1-l , 1-l ; 2 ; - \cot^2 \xi \right),
\\
\lim_{N \to \infty} C_{2l-1} (A^{(s)}, T^1_N) 
&= 0 \qquad (s \in \{m,f\}), 
\end{align*}
for $l=1,2, \ldots$ and $\xi \in (0,\pi/2).$ 
\label{kimarid1qwcr}
\end{cor}
\par
\
\par
(b) CRW case. 
\par
\
\par
One of the typical classes for $2 \times 2$ coin matrix $A^{(m)}$ (M-type) or $A^{(f)}$ (F-type) is as follows: 
\begin{align}
A^{(m)} 
=
\begin{bmatrix}
\cos^2 \xi & \sin^2 \xi  \\ 
\sin^2 \xi & \cos^2 \xi  
\end{bmatrix} 
, \qquad 
A^{(f)} 
=  
\begin{bmatrix}
\sin^2 \xi & \cos^2 \xi \\ 
\cos^2 \xi & \sin^2 \xi  
\end{bmatrix}
 \qquad ( \xi \in [0, 2\pi)).
\label{crwAMAF}
\end{align}
By Proposition \ref{kimarid1}, we have
\begin{cor} 
\begin{align*}
\overline{\zeta} \left(A^{(s)}, T^1_N, u \right)^{-1} 
&= \exp \left[ \frac{1}{N} \sum^{N-1}_{k =0} \log \left( F^{(s)} \left( \widetilde{k}, u \right)  \right) \right],
\\
\lim_{N \to \infty} \overline{\zeta} \left(A^{(s)}, T^1_N, u \right)^{-1} 
&= \exp \left[ \int_0^{2 \pi} \log \left( F^{(s)} \left( \theta, u \right) \right) \frac{d \theta}{2 \pi} \right],
\end{align*}
for $s \in \{m,f\}$, where
\begin{align*}
F^{(m)} \left( w, u \right) 
&= 1 - 2 \cos^2 \xi \cos w \cdot u + \cos (2 \xi) u^2,
\\
F^{(f)} \left( w, u \right) 
&= 1 - 2 \sin^2 \xi \cos w \cdot u - \cos (2 \xi) u^2.
\end{align*}
\label{kimarid1crw}
\end{cor}
Moreover, Lemma \ref{tukaeru} gives
\begin{cor} 
\begin{align*}
\lim_{N \to \infty} C_{2l} (A^{(m)}, T^1_N) 
&= 2 l \left(\cos \xi \right)^{4l} \sum_{m=1}^l \frac{1}{m} {l-1 \choose m-1}^2 \left( \tan^4 \xi \right)^{m}
\\
&= 2 l \left( \cos \xi \right)^{4(l-1)} (\sin^4 \xi) \ {}_2F_1 \left( 1-l , 1-l ; 2 ; \tan^4 \xi \right),
\\
\lim_{N \to \infty} C_{2l} (A^{(f)}, T^1_N) 
&= 2 l \left(\sin \xi \right)^{4l} \sum_{m=1}^l \frac{1}{m} {l-1 \choose m-1}^2 \left( \cot^4 \xi \right)^{m}
\\
&= 2 l \left(\sin \xi \right)^{4(l-1)} (\cos^4 \xi) \ {}_2F_1 \left( 1-l , 1-l ; 2 ; \cot^4 \xi \right),
\\
\lim_{N \to \infty} C_{2l-1} (A^{(s)}, T^1_N) 
&= 0 \qquad (s \in \{m,f\}), 
\end{align*}
for $l=1,2, \ldots$ and $\xi \in (0,\pi/2).$ 
\label{kimarid1crwcr}
\end{cor}
When we consider an extreme case $\xi = \pi/2$, i.e., 
\begin{align*}
A^{(m)} 
=
\begin{bmatrix}
0 & 1  \\ 
1 & 0   
\end{bmatrix} 
, \qquad 
A^{(f)} 
=  
\begin{bmatrix}
1 & 0 \\ 
0 & 1  
\end{bmatrix}
,
\end{align*}
then Corollary \ref{kimarid1crwcr} for $A^{(f)}$ (F-type) corresponds to results obtained by Komatsu et al. \cite{KomatsuEtAl}. Furthermore, if we take $2 \times 2$ coin matrix $A^{(m)}$ (M-type) or $A^{(f)}$ (F-type) as follows, 
\begin{align*}
A^{(m)} 
=
\begin{bmatrix}
c & d  \\ 
a & b   
\end{bmatrix} 
, \qquad 
A^{(f)} 
=  
\begin{bmatrix}
a & b \\ 
c & d  
\end{bmatrix}
,
\end{align*}
where $a+c=b+d=1$ and $a,b,c,d \in [0,1]$, then Corollary \ref{kimarid1crwcr} for $A^{(f)}$ (F-type) corresponds to results in Komatsu et al. \cite{KomatsuEtAl2020b}.  
\par
\
\par
(c) RW case. 
\par
\
\par
The RW is a special case of the CRW. The $2 \times 2$ coin matrix $A^{(m)}$ (M-type) or $A^{(f)}$ (F-type) for the RW is expressed in the following: 
\begin{align*}
A^{(m)} 
=
\begin{bmatrix}
\cos^2 \xi & \cos^2 \xi  \\ 
\sin^2 \xi & \sin^2 \xi  
\end{bmatrix} 
, \qquad 
A^{(f)} 
=  
\begin{bmatrix}
\sin^2 \xi & \sin^2 \xi \\ 
\cos^2 \xi & \cos^2 \xi  
\end{bmatrix}
 \qquad ( \xi \in [0, 2\pi)).
\end{align*}
Then a random walker moves at each step one unit to the left with probability $\cos^2 \xi$ or one unit to the right with probability $\sin^2 \xi$. From Proposition \ref{kimarid1}, we have
\begin{cor} 
\begin{align*}
\overline{\zeta} \left(A^{(s)}, T^1_N, u \right)^{-1} 
&= \exp \left[ \frac{1}{N} \sum^{N-1}_{k =0} \log \left( F^{(s)} \left( \widetilde{k}, u \right)  \right) \right],
\\
\lim_{N \to \infty} \overline{\zeta} \left(A^{(s)}, T^1_N, u \right)^{-1} 
&= \exp \left[ \int_0^{2 \pi} \log \left( F^{(s)} \left( \theta, u \right) \right) \frac{d \theta}{2 \pi} \right],
\end{align*}
for $s \in \{m,f\}$, where
\begin{align*}
F^{(m)} \left( w, u \right) 
&= 1 - \left( 2 i \cos^2 \xi \sin w + e^{-i w} \right) u,
\\
F^{(f)} \left( w, u \right) 
&= 1 + \left( 2 i \cos^2 \xi \sin w - e^{i w} \right) u.
\end{align*}
\label{kimarid1rw}
\end{cor}
Furthermore, by using Lemma \ref{tukaeru} and Eq. \eqref{atode}, we get
\begin{cor} 
\begin{align*}
\lim_{N \to \infty} C_{2l} (A^{(s)}, T^1_N) 
&= 2 l \left( \cos \xi \sin \xi \right)^{2l} \sum_{m=1}^l \frac{1}{m} {l-1 \choose m-1}^2 = \left( \cos \xi \sin \xi \right)^{2l} {2l \choose l},
\\
\lim_{N \to \infty} C_{2l-1} (A^{(s)}, T^1_N) 
&= 0, 
\end{align*}
for $s \in \{m,f\}, \ l=1,2, \ldots$ and $\xi \in (0, \pi/2).$ 
\label{kimarid1crwrw}
\end{cor}
We should note that $\lim_{N \to \infty} C_{2l} (A^{(s)}, T^1_N)$ is nothing but the return probability of the RW at time $2l$. However, the corresponding value for the CRW in Corollary \ref{kimarid1crwcr} is not the same as the return probability of the CRW at time $2l$. In general, such a correspondence is limited to the case of the RW.

When $\xi = \pi/4$ (symmetric RW), then $2 \times 2$ coin matrices $A^{(m)}$ (M-type) and $A^{(f)}$ (F-type) become 
\begin{align*}
A^{(m)} 
= A^{(f)} = \frac{1}{2}
\begin{bmatrix}
1 & 1  \\ 
1 & 1  
\end{bmatrix} 
.
\end{align*}
Therefore we have
\begin{cor}
\begin{align*}
\overline{\zeta} \left(A^{(s)}, T^1_N, u \right)^{-1} 
& = \exp \left[ \frac{1}{N} \sum^{N-1}_{ k =0} \log \left( 1 -  \cos \widetilde{k} \cdot u \right) \right],
\\
\lim_{N \to \infty} \overline{\zeta} \left(A^{(s)}, T^1_N, u \right)^{-1}
&= \exp \left[ \int_0^{2 \pi} \log \left( 1 -  \cos \theta \cdot u \right) \frac{d \theta}{2 \pi} \right],
\\
\lim_{N \to \infty} C_{2l} (A^{(s)}, T^1_N) 
&= \left( \frac{1}{2} \right)^{2l} {2l \choose l},
\\
\lim_{N \to \infty} C_{2l-1} (A^{(s)}, T^1_N) 
&= 0, 
\end{align*}
for $s \in \{m,f\}$ and $l=1,2, \ldots$.
\label{kimarid1symrw}
\end{cor}

Next we consider a generalized version of our walk on $T_N^1$, whose $3 \times 3$ coin matrix $A^{(m)}$ (M-type) or $A^{(f)}$ (F-type) is defined as follows: 
\begin{align*}
A^{(m)} 
=
\begin{bmatrix}
a_{11} & a_{12} & a_{13} \\ 
a_{21} & a_{22} & a_{23} \\ 
a_{31} & a_{32} & a_{33} \\ 
\end{bmatrix} 
, \qquad 
A^{(f)} 
=  
\begin{bmatrix}
a_{31} & a_{32} & a_{33} \\ 
a_{21} & a_{22} & a_{23} \\ 
a_{11} & a_{12} & a_{13} 
\end{bmatrix}
.
\end{align*}
In this case, we take the projections $\{ P_0, P_1, P_2 \}$ by 
\begin{align*}
P_{1} 
=
\begin{bmatrix}
1 & 0 & 0 \\ 
0 & 0 & 0 \\
0 & 0 & 0
\end{bmatrix}
, \qquad 
P_{0} 
=
\begin{bmatrix}
0 & 0 & 0 \\ 
0 & 1 & 0 \\
0 & 0 & 0 
\end{bmatrix}
, \qquad 
P_{2} 
=
\begin{bmatrix}
0 & 0 & 0 \\ 
0 & 0 & 0 \\
0 & 0 & 1 
\end{bmatrix}
. 
\end{align*}
Similarly, we define a $3 N \times 3 N$ matrix
\begin{align*}
M_A = P_{1} A \tau^{-1} + P_{0} A + P_{2} A \tau.
\end{align*}
Then the walker moves at each step one unit to the left with $P_{1}$ or one unit to the right with $P_{2}$ or stays at each step with $P_{0}$. Thus we get
\begin{align*}                          
\det \Big( I_{3} - u \widehat{M}_A (\widetilde{\kvec}) \Big) = \det \left(I_{3} - u \times \Big( e^{i \widetilde{k}} P_{1} A + P_0 A + e^{-i \widetilde{k}} P_{2} A \Big) \right). 
\end{align*}
A typical example is the three-state Grover walk whose $3 \times 3$ coin matrix $A^{(m)}$ (M-type) or $A^{(f)}$ (F-type) is defined by 
\begin{align}
A^{(m)} 
= \frac{1}{3}
\begin{bmatrix}
-1 & 2 & 2 \\ 
2 & -1 & 2 \\ 
2 & 2 & -1 \\ 
\end{bmatrix} 
, \qquad 
A^{(f)} 
= 
\frac{1}{3}
\begin{bmatrix}
2 & 2 & -1 \\ 
2 & -1 & 2 \\ 
-1 & 2 & 2 
\end{bmatrix}
.
\label{3stateA}
\end{align}
Here $A^{(m)}$ is the $3 \times 3$ Grover matrix. In general, the $n \times n$ {\em Grover matrix} $G^{(n)} =[G^{(n)}_{ab}]_{a,b=1,2,\ldots,d}$ is defined by
\begin{align*}
G^{(n)}_{aa} = \frac{2}{n} - 1, \qquad G^{(n)}_{ab} = \frac{2}{n} \quad (a \not= b).
\end{align*}
We should note that $G^{(n)}$ is unitary. The walk defined by the Grover matrix is called the {\em Grover walk}. Then, $A^{(m)}$ and $A^{(f)}$ in Eq. \eqref{3stateA} are unitary, so the walks determined by them become QWs. In a similar fashion, we obtain 
\begin{cor} 
\begin{align}
\overline{\zeta} \left(A^{(s)}, T^1_N, u \right)^{-1} 
&= (1+(-1)^{\gamma(s)} u) \exp \left[ \frac{1}{N} \sum^{N-1}_{k =0} \log \left( F^{(s)} \left( \widetilde{k}, u \right)  \right) \right],
\label{kimarid13qwf}
\\
\lim_{N \to \infty} \overline{\zeta} \left(A^{(s)}, T^1_N, u \right)^{-1} 
&= (1+(-1)^{\gamma(s)} u) \exp \left[ \int_0^{2 \pi} \log \left( F^{(s)} \left( \theta, u \right) \right) \frac{d \theta}{2 \pi} \right],
\label{kimarid13qwnf}
\end{align}
for $s \in \{m,f\}$, where $\gamma(s)=1$ for $s=m$, $\gamma(s)=0$ for $s=f$, and 
\begin{align*}
F^{(s)} \left( w, u \right) 
= 1 - \frac{(-1)^{\gamma(s)} \cdot \ 2}{3}  \left( 1 + 2 \cos w + \gamma(s) (1 - \cos w ) \right) u + u^2.
\end{align*}
\label{kimarid13qw}
\end{cor}
Remark that the leading factor $(1+(-1)^{\gamma(s)} u)$ of the right-hand side of Eqs. \eqref{kimarid13qwf} and \eqref{kimarid13qwnf} corresponds to localization of the three-state Grover walk on $\ZM$ (see Konno \cite{Konno2008}, for example). Localization means that limsup for time $n$ of the probability that the walker returns to the starting location at time $n$ is positive. 

In the final part of this section, we consider the OQRW on $T_N^1$, whose dynamics is defined by $4 \times 4$ matrix determined by two $2 \times 2$ matrices $B$ and $C$. Here $B$ and $C$ satisfy
\begin{align*}
B^{\ast} B + C^{\ast} C = I_2,
\end{align*}
where $\ast$ means the adjoint operator. The OQRW was introduced by Attal et al. \cite{AttalEtAl2012b, AttalEtAl2012a}. Put
\begin{align*}
B
=
\begin{bmatrix}
b_{11} & b_{12}  \\ 
b_{21} & b_{22}   
\end{bmatrix} 
, \qquad 
C 
=
\begin{bmatrix}
c_{11} & c_{12}  \\ 
c_{21} & c_{22}  
\end{bmatrix}
.
\end{align*}
Let $2 \times 2$ matrix $\rho_{n}(x)$ denote the state at time $n \in \mathbb{Z}_{\ge}$ and location $x \in T^1_N$ for OQRW. The evolution of OQRW is determined by
\begin{align}
\rho_{n+1}(x) = B \rho_{n}(x+1) B^{\ast} + C \rho_{n}(x-1) C^{\ast}.
\label{oqrwEVO}
\end{align}
The measure at time $n \in \mathbb{Z}_{\ge}$ and location $x \in T^1_N$ is defined by
\begin{align*}
\mu_n (x) = {\rm Tr} \left( \rho_{n}(x) \right).
\end{align*}
From now on, we consider $2 \times 2$ matrix $\rho_{n}(x)$ as the following four-dimensional vector:
\begin{align*}
\rho_{n}(x)=
\begin{bmatrix}
\rho^{11}_{n}(x) \\ \rho^{12}_{n}(x) \\ \rho^{21}_{n}(x) \\ \rho^{22}_{n}(x) 
\end{bmatrix} 
\in \mathbb{C}^{4}.
\end{align*}
In this setting, Eq. \eqref{oqrwEVO} can be rewritten as
\begin{align}
\rho_{n+1}(x) = \widetilde{P}_B \ \rho_{n}(x+1) + \widetilde{P}_C \ \rho_{n}(x-1),
\label{oqrwnaradewa}
\end{align}
where
\begin{align}
\widetilde{P}_B
&= B \otimes \bar{B}
=
\begin{bmatrix}
|b_{11}|^2 & b_{11} \overline{b_{12}} & \overline{b_{11}} b_{12} & |b_{12}|^2 \\b_{11} \overline{b_{21}} & b_{11} \overline{b_{22}} & b_{12} \overline{b_{21}} & b_{12} \overline{b_{22}} \\
\overline{b_{11}} b_{21} & \overline{b_{12}} b_{21} & \overline{b_{11}} b_{22} & \overline{b_{12}} b_{22} \\
|b_{21}|^2 & b_{21} \overline{b_{22}} & \overline{b_{21}} b_{22} & |b_{22}|^2 
\end{bmatrix} 
, \qquad 
\label{BC1}
\\
\widetilde{P}_C
&= C \otimes \bar{C}
=
\begin{bmatrix}
|c_{11}|^2 & c_{11} \overline{c_{12}} & \overline{c_{11}} c_{12} & |c_{12}|^2 \\c_{11} \overline{c_{21}} & c_{11} \overline{c_{22}} & c_{12} \overline{c_{21}} & c_{12} \overline{c_{22}} \\
\overline{c_{11}} c_{21} & \overline{c_{12}} c_{21} & \overline{c_{11}} c_{22} & \overline{c_{12}} c_{22} \\
|c_{21}|^2 & c_{21} \overline{c_{22}} & \overline{c_{21}} c_{22} & |c_{22}|^2 
\end{bmatrix}
.
\label{BC2}
\end{align}
Here each component of $\bar{A}$ is the complex conjugate of that of $A$ for a matrix $A$. The Fourier transform of $\rho_{n}(x)$, denoted by $\widehat{\rho}_{n}(\widetilde{k})$, is defined by the sum
\begin{align*}
\widehat{\rho}_{n}(\widetilde{k}) = \frac{1}{\sqrt{N}} \sum_{x \in \mathbb{K}_N} e^{- i \widetilde{k} x} \rho_{n}(x).
\end{align*}
Then we have
\begin{align*}
\widehat{\rho}_{n+1}(\widetilde{k}) = \widehat{M}_{B,C} (\widetilde{k}) \widehat{\rho}_{n}(\widetilde{k}),
\end{align*}
where
\begin{align*}                          
\widehat{M}_{B,C}(\widetilde{k})= e^{i \widetilde{k}} \widetilde{P}_B + e^{-i \widetilde{k}} \widetilde{P}_C. 
\end{align*}
Note that $\widehat{M}_{B,C} (\widetilde{k})$ is a counterpart of $\widehat{M}_A (\widetilde{k})$ in Eqs. \eqref{jinji01} and \eqref{jinji02}, that is, 
\begin{align}
\widehat{M}_{A^{(s)}} (\widetilde{k})
= e^{i \widetilde{k}} P_{1} A^{(s)} + e^{-i \widetilde{k}} P_{2} A^{(s)} \quad (s \in \{ m, f \}). 
\label{senbatsu}
\end{align}

One of the typical model (see \cite{KonnoYoo}, for example) is given by
\begin{align*}
B
=
\frac{1}{\sqrt{3}}
\begin{bmatrix}
1 & 1  \\ 
0 & 1   
\end{bmatrix} 
, \qquad 
C 
=
\frac{1}{\sqrt{3}}  
\begin{bmatrix}
1 & 0 \\ 
-1 & 1  
\end{bmatrix}
.
\end{align*}
In this case, Eqs. \eqref{BC1} and \eqref{BC2} imply
\begin{align*}
\widetilde{P}_B
=
\frac{1}{3}
\begin{bmatrix}
1 & 1 & 1 & 1 \\ 
0 & 1 & 0 & 1 \\ 
0 & 0 & 1 & 1 \\ 
0 & 0 & 0 & 1 \\  
\end{bmatrix}
, \quad
\widetilde{P}_C
=
\frac{1}{3}
\begin{bmatrix}
1 & 0 & 0 & 0 \\ 
-1 & 1 & 0 & 0 \\ 
-1 & 0 & 1 & 0 \\ 
1 & -1 & -1 & 1 \\  
\end{bmatrix}
.
\end{align*}
By using these, we get
\begin{align*}
\widehat{M}_{B,C} (\widetilde{k})
=
\frac{1}{3}  
\begin{bmatrix}
2 \cos \widetilde{k} & e^{i \widetilde{k}} & e^{i \widetilde{k}} & e^{i \widetilde{k}} \\ 
- e^{-i \widetilde{k}} & 2 \cos \widetilde{k} & 0 & e^{i \widetilde{k}} \\ 
- e^{-i \widetilde{k}} & 0 & 2 \cos \widetilde{k} & e^{i \widetilde{k}} \\ 
e^{-i \widetilde{k}} & - e^{-i \widetilde{k}} & - e^{-i \widetilde{k}} & 2 \cos \widetilde{k}  
\end{bmatrix}
.
\end{align*}
Similarly, by computing $\det \Big( I_{4} - u \widehat{M}_{B,C} (\widetilde{k}) \Big)$, we have
\begin{cor} 
\begin{align*}
\overline{\zeta} \left(A, T^1_N, u \right)^{-1} 
&= \exp \left[ \frac{1}{N} \sum^{N-1}_{k =0} \log \left( F \left( \widetilde{k}, u \right)  \right) \right],
\\
\lim_{N \to \infty} \overline{\zeta} \left(A, T^1_N, u \right)^{-1} 
&=  \exp \left[ \int_0^{2 \pi} \log \left( F \left( \theta, u \right) \right) \frac{d \theta}{2 \pi} \right],
\end{align*}
where
\begin{align*}
F \left( w, u \right) 
&= 1 - \frac{8 \cos w}{3} \ u + \frac{8 \cos^2 w +1}{3} \ u^2 
\\
& \qquad \qquad - \frac{16}{27} \cos w \left( 2 \cos^2 w + 1 \right) u^3 +  \frac{4}{81} \cos^2 w \left( 4 \cos^2 w + 5 \right) u^4.
\end{align*}
\label{kimarid1oqrw}
\end{cor}

Finally, we deals with a relation between OQRW and CRW. To do so, we take 
\begin{align*}
B
=
\begin{bmatrix}
b_{11} & 0  \\ 
b_{21} & 0   
\end{bmatrix} 
, \qquad 
C 
=
\begin{bmatrix}
0 & c_{12}  \\ 
0 & c_{22}  
\end{bmatrix}
.
\end{align*}
Then assumption $B^{\ast} B + C^{\ast} C = I_2$ gives 
\begin{align*}
|b_{11}|^2 + |b_{21}|^2 = |c_{12}|^2 + |c_{22}|^2 = 1.
\end{align*}
Moreover, by Eqs. \eqref{BC1} and \eqref{BC2}, we have 
\begin{align*}
\widetilde{P}_B
=
\begin{bmatrix}
|b_{11}|^2 & 0 & 0 & 0 \\
b_{11} \overline{b_{21}} & 0 & 0 & 0 \\
\overline{b_{11}} b_{21} & 0 & 0 & 0 \\
|b_{21}|^2 & 0 & 0 & 0
\end{bmatrix} 
, \qquad 
\widetilde{P}_C
=
\begin{bmatrix}
0 & 0 & 0 & |c_{12}|^2 \\
0 & 0 & 0 & c_{12} \overline{c_{22}} \\
0 & 0 & 0 & \overline{c_{12}} c_{22} & \\
0 & 0 & 0 & |c_{22}|^2
\end{bmatrix}
.
\end{align*}
For the state of this model, we can reduce from four-dimensional vector to two-dimensional vector denoted by
\begin{align*}
\rho_{n}^{(r)} (x)=
\begin{bmatrix}
\rho^{11}_{n}(x)  \\ \rho^{22}_{n}(x) 
\end{bmatrix} 
\in \mathbb{C}^{2}.
\end{align*}
It follows from Eq. \eqref{oqrwnaradewa} that 
\begin{align*}
\rho_{n+1}^{(r)}(x) = \widetilde{P}_B^{(r)} \ \rho_{n}^{(r)}(x+1) + \widetilde{P}_C^{(r)} \ \rho_{n}^{(r)}(x-1),
\end{align*}
where
\begin{align*}
\widetilde{P}_B^{(r)}
=
\begin{bmatrix}
|b_{11}|^2 & 0  \\
|b_{21}|^2 & 0 
\end{bmatrix} 
, \qquad 
\widetilde{P}_C^{(r)}
=
\begin{bmatrix}
0 & |c_{12}|^2 \\
0 & |c_{22}|^2
\end{bmatrix}
.
\end{align*}
Therefore we have
\begin{align*}
\widehat{\rho}_{n+1}^{(r)}(\widetilde{k}) = \widehat{M}_{B,C}^{(r)} (\widetilde{k}) \widehat{\rho}_{n}^{(r)}(\widetilde{k}),
\end{align*}
where
\begin{align}                          
\widehat{M}_{B,C}^{(r)}(\widetilde{k})= e^{i \widetilde{k}} \widetilde{P}_B^{(r)} + e^{-i \widetilde{k}} \widetilde{P}_C^{(r)}. 
\label{ichikoro}
\end{align}
Here we introduce the corresponding coin matrix $A$ for the CRW as follows:
\begin{align*}
A 
=
\begin{bmatrix}
|b_{11}|^2 & |c_{12}|^2 \\
|b_{21}|^2 & |c_{22}|^2
\end{bmatrix}
\end{align*}
with $|b_{11}|^2 + |b_{21}|^2 = |c_{12}|^2 + |c_{22}|^2 = 1$. Remark that $A=\widetilde{P}_B^{(r)}+\widetilde{P}_C^{(r)}$. In this case, we see
\begin{align*}
\widetilde{P}_B^{(r)} = A P_1, \quad \widetilde{P}_C^{(r)} = A P_2,
\end{align*}
where
\begin{align*}
P_1
=
\begin{bmatrix}
1 & 0  \\
0 & 0 
\end{bmatrix} 
, \qquad 
P_2
=
\begin{bmatrix}
0 & 0  \\
0 & 1 
\end{bmatrix}
.
\end{align*}
Thus Eq. \eqref{ichikoro} becomes
\begin{align}                          
\widehat{M}_{B,C}^{(r)}(\widetilde{k})= e^{i \widetilde{k}} A P_1 + e^{-i \widetilde{k}} A P_2. 
\label{ichikoroda01}
\end{align}
We recall the expression in Eq. \eqref{senbatsu}, that is,
\begin{align}
\widehat{M}_{A^{(s)}} (\widetilde{k})
= e^{i \widetilde{k}} P_{1} A^{(s)} + e^{-i \widetilde{k}} P_{2} A^{(s)} \quad (s \in \{ m, f \}). 
\label{ichikoroda02}
\end{align}
So this walker defined by Eq. \eqref{ichikoroda01} moves at each step one unit to the left with $AP_1$ or to the right with $AP_2$. On the other hand, the previous walker defined by Eq. \eqref{ichikoroda02} moves at each step one unit to the left with $P_1 A^{(s)}$ or to the right with $P_2  A^{(s)}$ for $s \in \{ m,f \}$. This is the connection between OQRW and CRW we wanted to mention.

In this model, following Eq. \eqref{crwAMAF}, we put
\begin{align*}
A 
=
\begin{bmatrix}
|b_{11}|^2 & |c_{12}|^2 \\
|b_{21}|^2 & |c_{22}|^2
\end{bmatrix}
=
\begin{bmatrix}
\cos^2 \xi & \sin^2 \xi  \\ 
\sin^2 \xi & \cos^2 \xi  
\end{bmatrix} 
(=A^{(m)}). 
\end{align*}
Then Eq. \eqref{ichikoroda01} implies
\begin{align*}                          
\widehat{M}_{B,C}^{(r)}(\widetilde{k}) 
=
\begin{bmatrix}
e^{i \widetilde{k}} \cos^2 \xi & e^{-i \widetilde{k}} \sin^2 \xi  \\ 
e^{i \widetilde{k}} \sin^2 \xi & e^{-i \widetilde{k}} \cos^2 \xi  
\end{bmatrix}
. 
\end{align*}
In a similar way, we obtain
\begin{cor} 
\begin{align*}
\overline{\zeta} \left(A, T^1_N, u \right)^{-1} 
&= \exp \left[ \frac{1}{N} \sum^{N-1}_{k =0} \log \left( F \left( \widetilde{k}, u \right)  \right) \right],
\\
\lim_{N \to \infty} \overline{\zeta} \left(A, T^1_N, u \right)^{-1} 
&= \exp \left[ \int_0^{2 \pi} \log \left( F \left( \theta, u \right) \right) \frac{d \theta}{2 \pi} \right],
\end{align*}
where
\begin{align*}
F \left( w, u \right) 
= 1 - 2 \cos^2 \xi \cos w \cdot u + \cos (2 \xi) u^2.
\end{align*}
\label{kimarid1crwoqrw}
\end{cor}
We should note that this result is the same as $A^{(m)}$ case in Corollary \ref{kimarid1crw}. Furthermore, for this model, we change Lemma \ref{tukaerupre} as follows:
\begin{align*}
Q_1 = A P_1, \quad Q_2 = A P_2, \quad Q_3 = A P_2 \sigma, \quad Q_4 = A P_1 \sigma, 
\end{align*}
that is,
\begin{align*}
Q_1
=
\begin{bmatrix}
a_{11} & 0 \\ 
a_{21} & 0 
\end{bmatrix} 
, \qquad 
Q_2
=
\begin{bmatrix}
0 & a_{12}  \\
0 & a_{22} 
\end{bmatrix} 
, \qquad
Q_3
=
\begin{bmatrix}
a_{12} & 0 \\
a_{22} & 0 
\end{bmatrix} 
, \qquad 
Q_4
=
\begin{bmatrix}
0 & a_{11} \\ 
0 & a_{21}
\end{bmatrix} 
.
\end{align*}
Noting that Eq. \eqref{matsuyama} is also correct for this case, we can confirm that Lemma \ref{tukaeru} holds. Therefore we have the same result as $A^{(m)}$ case in Corollary \ref{kimarid1crwcr}.

\section{Example \label{secex}}
This secton is devoted to specific examples of QWs and CRWs with M-type in the previous section for a better understanding of Walk/Zeta Correspondence. Other models can be considered in a similar fashion.   
\par
\
\par
(a) QW case. 
\par
\
\par
We deal with the Hadamard walk ($\xi = \pi/4$) with M-type whose $2 \times 2$ coin matrix $A^{(m)}$ is given by
\begin{align*}
A^{(m)} 
=
\begin{bmatrix}
\cos (\pi/4) & \sin (\pi/4)  \\ 
\sin (\pi/4) & - \cos (\pi/4)  
\end{bmatrix} 
=
\frac{1}{\sqrt{2}}
\begin{bmatrix}
1 & 1 \\ 
1 & -1  
\end{bmatrix}
.
\end{align*}
From Corollary \ref{kimarid1qwcr}, we obtain
\begin{align*}
C_{2l} (0) 
&= \lim_{N \to \infty} C_{2l} (A^{(m)}, T^1_N) 
\\
&= l \left(- \frac{1}{2} \right)^{l-1}  \ {}_2F_1 \left( 1-l , 1-l ; 2 ; - 1 \right)
\\
&= l \left(- \frac{1}{2} \right)^{l-1} \sum_{m=1}^l \frac{1}{m} \left( - 1 \right)^{m-1}{l-1 \choose m-1}^2.
\end{align*}
For example, by using these, we get 
\begin{align*}
C_{2} (0) = 1, \qquad C_{4} (0) = - \frac{1}{2}. 
\end{align*}
It follows from Theorem \ref{satosan10thm} that
\begin{align*}
C_{2l} (0) = {\rm Tr} \left( \Phi_{2l} ^{(\infty)} (0) \right).
\end{align*}
Next we compute $2 \times 2$ matrix $\Phi_{2l} ^{(\infty)} (0)$ by using notations in Lemma \ref{tukaerupre} as follows.
\begin{align*}
\Phi_{2} ^{(\infty)} (0) 
&= Q_1 Q_2 + Q_2 Q_1 = \frac{1}{2}
\begin{bmatrix}
1 & -1 \\ 
1 & 1  
\end{bmatrix}
,
\\
\Phi_{4} ^{(\infty)} (0) 
&= (Q_1 Q_2 + Q_2 Q_1)^2 + Q_1^2 Q_2^2 + Q_2^2 Q_1^2 = \frac{1}{4}
\begin{bmatrix}
-1 & -1 \\ 
1 & -1  
\end{bmatrix}
,
\end{align*}
where
\begin{align*}
Q_1 = \frac{1}{\sqrt{2}}
\begin{bmatrix}
1 & 1 \\ 
0 & 0  
\end{bmatrix}
, \qquad 
Q_2 = \frac{1}{\sqrt{2}}
\begin{bmatrix}
0 & 0 \\ 
1 & -1  
\end{bmatrix}
.
\end{align*}
So we have 
\begin{align*}
C_{2} (0) = {\rm Tr} \left( \Phi_{2} ^{(\infty)} (0) \right) = 1, \qquad C_{4} (0) = {\rm Tr} \left( \Phi_{4} ^{(\infty)} (0) \right) = - \frac{1}{2}. 
\end{align*}
Therefore we can confirm that Corollary \ref{kimarid1qwcr} holds for $l=1$ and $l=2$.

On the other hand, the return probability $\mu_{2l} (0)$ for the Hadamard walk starting from the origin at time $2 l$ depends on the initial state $\varphi = {}^T [\alpha, \beta] \in \CM^2$ with $|\alpha|^2 + |\beta|^2 =1$, where $T$ is the transposed operator. Then 
\begin{align*}
\mu_{2 l} (0) = \| \Phi_{2 l} ^{(\infty)} (0) \varphi \|_{\mathbb{C}^{2}}^2.
\end{align*}
For example, if we take $\varphi = {}^T [1/\sqrt{2}, i/\sqrt{2}]$, then we obtain
\begin{align*}
\mu_{0} (0) = 1, \quad \mu_{2} (0) = \frac{1}{2}, \quad \mu_{4} (0) = \mu_{6} (0) = \frac{1}{8}, \quad \ldots.
\end{align*}
Remark that it is known in Konno \cite{Konno2010b} that
\begin{align*}
\mu_{4m} (0) = \mu_{4m+2} (0) = \frac{1}{2^{4m+1}} {2m \choose m}^2 \qquad (m \ge 1).
\end{align*}
Therefore we see that in general $\mu_{2 l} (0)$ is not the same as $C_{2 l} (0)$.

\par
\
\par
(b) CRW case. 
\par
\
\par
As in the case of the Hadamard walk, we consider a CRW ($\xi = \pi/6$) with M-type whose $2 \times 2$ coin matrix $A^{(m)}$ is given by
\begin{align*}
A^{(m)} 
=
\begin{bmatrix}
\cos^2 (\pi/6) & \sin^2 (\pi/6)  \\ 
\sin^2 (\pi/6) & \cos^2 (\pi/6)  
\end{bmatrix} 
=
\frac{1}{4}
\begin{bmatrix}
3 & 1 \\ 
1 & 3  
\end{bmatrix}
.
\end{align*}
By Corollary \ref{tukaeru}, we get
\begin{align*}
C_{2l} (0) 
&= \lim_{N \to \infty} C_{2l} (A^{(m)}, T^1_N) 
\\
&= \frac{3^{2(l-1)} \ l}{2^{4l-1}} \ {}_2F_1 \left( 1-l , 1-l ; 2 ; 1/9 \right)
\\
&= \frac{3^{2(l-1)} \ l}{2^{4l-1}} \sum_{m=1}^l \frac{1}{m} \left( \frac{1}{9} \right)^{m-1}{l-1 \choose m-1}^2.
\end{align*}
For example, from using these, we have 
\begin{align*}
C_{2} (0) = \frac{1}{8}, \qquad C_{4} (0) = \frac{19}{128} = 0.1484 \ldots. 
\end{align*}
Next we calculate $2 \times 2$ matrix $\Phi_{2l} ^{(\infty)} (0)$ by using notations in Lemma \ref{tukaerupre} like the Hadamard walk case:
\begin{align*}
\Phi_{2} ^{(\infty)} (0) 
&= Q_1 Q_2 + Q_2 Q_1 = \frac{1}{16}
\begin{bmatrix}
1 & 3 \\ 
3 & 1  
\end{bmatrix}
,
\\
\Phi_{4} ^{(\infty)} (0) 
&= (Q_1 Q_2 + Q_2 Q_1)^2 + Q_1^2 Q_2^2 + Q_2^2 Q_1^2 = \left( \frac{1}{16} \right)^2
\begin{bmatrix}
19 & 33 \\ 
33 & 19  
\end{bmatrix}
,
\end{align*}
where
\begin{align*}
Q_1 = \frac{1}{4}
\begin{bmatrix}
3 & 1 \\ 
0 & 0  
\end{bmatrix}
, \qquad 
Q_2 = \frac{1}{4}
\begin{bmatrix}
0 & 0 \\ 
1 & 3  
\end{bmatrix}
.
\end{align*}
Thus we get
\begin{align*}
C_{2} (0) = {\rm Tr} \left( \Phi_{2} ^{(\infty)} (0) \right) = \frac{1}{8}, \qquad C_{4} (0) = {\rm Tr} \left( \Phi_{4} ^{(\infty)} (0) \right) = \frac{19}{128}. 
\end{align*}
Therefore we can confirm that Corollary \ref{tukaeru} is valid for $l=1$ and $l=2$.

On the other hand, the return probability $\mu_{2l} (0)$ for the CRW starting from the origin at time $2 l$ depends on the initial state $\varphi = {}^T [\alpha, \beta] \in [0,1]^2$ with $\alpha + \beta =1$. Then 
\begin{align*}
\mu_{2 l} (0) = \| \Phi_{2 l} ^{(\infty)} (0) \varphi \|_{\mathbb{R}^{2}}^1.
\end{align*}
For instance, if we take $\varphi = {}^T [1/2, 1/2]$, then we have
\begin{align*}
\mu_{0} (0) = 1, \quad \mu_{2} (0) = \frac{1}{4}, \quad \mu_{4} (0) = \frac{13}{64} = 0.2031, \quad \ldots.
\end{align*}
Thus we see that in general $\mu_{2 l} (0)$ is different from $C_{2 l} (0)$ as in the case of the Hadamard walk.

\section{Two-Dimensional Case \label{sec05}} 
This section treats walks on the two-dimensional torus $T^2_N$ whose $4 \times 4$ coin matrix $A^{(m)}$ (M-type) or $A^{(f)}$ (F-type) as follows: 
\begin{align*}
A^{(m)} 
=
\begin{bmatrix}
a_{11} & a_{12} & a_{13} & a_{14} \\ 
a_{21} & a_{22} & a_{23} & a_{24} \\ 
a_{31} & a_{32} & a_{33} & a_{34} \\ 
a_{41} & a_{42} & a_{43} & a_{44} 
\end{bmatrix} 
, \qquad 
A^{(f)} 
=  
\begin{bmatrix}
a_{21} & a_{22} & a_{23} & a_{24} \\
a_{11} & a_{12} & a_{13} & a_{14} \\ 
a_{41} & a_{42} & a_{43} & a_{44} \\
a_{31} & a_{32} & a_{33} & a_{34}  
\end{bmatrix}
,
\end{align*}
since
\begin{align*}
A^{(f)} = \left( I_2 \otimes \sigma \right) A^{(m)}.
\end{align*}
In this case, we take 
\begin{align*}
P_{1} 
=
\begin{bmatrix}
1 & 0 & 0 & 0 \\ 
0 & 0 & 0 & 0 \\ 
0 & 0 & 0 & 0 \\ 
0 & 0 & 0 & 0  
\end{bmatrix}
, \quad 
P_{2} 
=
\begin{bmatrix}
0 & 0 & 0 & 0 \\ 
0 & 1 & 0 & 0 \\ 
0 & 0 & 0 & 0 \\ 
0 & 0 & 0 & 0  
\end{bmatrix}
, \quad 
P_{3} 
=
\begin{bmatrix}
0 & 0 & 0 & 0 \\ 
0 & 0 & 0 & 0 \\ 
0 & 0 & 1 & 0 \\ 
0 & 0 & 0 & 0  
\end{bmatrix}
, \quad 
P_{4} 
=
\begin{bmatrix}
0 & 0 & 0 & 0 \\ 
0 & 0 & 0 & 0 \\ 
0 & 0 & 0 & 0 \\ 
0 & 0 & 0 & 1  
\end{bmatrix}
.
\end{align*}
Similarly, we define a $4 N^2 \times 4 N^2$ matrix
\begin{align*}
M_A = P_{1} A \tau_1^{-1} + P_{2} A \tau_1 +  P_{3} A \tau_2^{-1} + P_{4} A \tau_2.
\end{align*}
Then the walker moves at each step one unit to the left with $P_{1}$ or to the right with $P_{2}$ or to the down with $P_{3}$ or to the up with $P_{4}$. Thus we get
\begin{align*}                          
\det \Big( I_{4} - u \widehat{M}_A (\widetilde{\kvec}) \Big) = \det \left(I_{4} - u \times \Big( e^{i \widetilde{k}_1} P_{1} A + e^{-i \widetilde{k}_1} P_{2} A + e^{i \widetilde{k}_2} P_{3} A + e^{-i \widetilde{k}_2} P_{4} A \Big) \right). 
\end{align*}
A typical example is the two-dimensional Grover walk whose $4 \times 4$ coin matrix $A^{(m)}$ (M-type) or $A^{(f)}$ (F-type) is defined by 
\begin{align*}
A^{(m)} 
= \frac{1}{2}
\begin{bmatrix}
-1 & 1 & 1 & 1 \\ 
1 & -1 & 1 & 1 \\ 
1 & 1 & -1 & 1 \\ 
1 & 1 & 1 & -1 
\end{bmatrix} 
, \qquad 
A^{(f)} 
= 
\frac{1}{2}
\begin{bmatrix}
1 & -1 & 1 & 1 \\ 
-1 & 1 & 1 & 1 \\ 
1 & 1 & 1 & -1 \\ 
1 & 1 & -1 & 1 
\end{bmatrix}
.
\end{align*}
Here $A^{(m)}$ is the $4 \times 4$ Grover matrix. Then we have
\begin{cor}
\begin{align}
\overline{\zeta} \left(A^{(s)}, T^2_N, u \right)^{-1} 
& = (1-u^2) \exp \left[ \frac{1}{N^2} \sum^{N-1}_{k_1 =0} \sum^{N-1}_{k_2 =0} \log \left( F^{(s)} \left( \widetilde{k}_1, \widetilde{k}_{2}, u \right) \right) \right],
\label{manbou001}
\\
\lim_{N \to \infty} \overline{\zeta} \left(A^{(s)}, T^2_N, u \right)^{-1} 
& = (1-u^2) \exp \left[ \int_0^{2 \pi} \int_0^{2 \pi} \log \left( F^{(s)} \left( \theta_1, \theta_2, u \right) \right) \frac{d \theta_1}{2 \pi} \frac{d \theta_2}{2 \pi} \right],
\label{manbou002}
\end{align}
for $s \in \{m,f\}$, where
\begin{align*}
F^{(m)} \left( w_1, w_2, u \right) 
&= 1 + \left( \cos w_1 + \cos w_{2} \right) u + u^2,
\\
F^{(f)} \left( w_1, w_2, u \right) 
&= 1 - \left( \cos w_1 + \cos w_{2} \right) u + u^2.
\end{align*}
\label{kimarid2grover}
\end{cor}
Note that the leading factor $(1- u^2)$ of the right-hand side of Eqs. \eqref{manbou001} and \eqref{manbou002} corresponds to localization of the four-state Grover walk on $\ZM^2$ (see \cite{KomatsuKonno}, for example). Moreover, $F^{(s)} \left( w_1, w_2, u \right)$ for $s \in \{m,f\}$ are the same as the corresponding results in Asano et al. \cite{AsanoEtAl2019}.

Another typical example is the two-dimensional Fourier walk whose $4 \times 4$ coin matrix $A^{(m)}$ (M-type) or $A^{(f)}$ (F-type) is determined by 
\begin{align*}
A^{(m)} 
= \frac{1}{2}
\begin{bmatrix}
1 & 1 & 1 & 1 \\ 
1 & i & -1 & -i \\ 
1 & -1 & 1 & -1 \\ 
1 & -i & -1 & i 
\end{bmatrix} 
, \qquad 
A^{(f)} 
= 
\frac{1}{2}
\begin{bmatrix}
1 & i & -1 & -i \\
1 & 1 & 1 & 1 \\ 
1 & -i & -1 & i \\
1 & -1 & 1 & -1 
\end{bmatrix}
.
\end{align*}
Here $A^{(m)}$ is the $4 \times 4$ Fourier matrix. In general, the $n \times n$ {\em Fourier matrix} $F^{(n)} =[F^{(n)}_{ab}]_{a,b=1,2,\ldots,d}$ is defined by
\begin{align*}
F^{(n)}_{ab} = \frac{1}{\sqrt{n}} \ \omega_n^{(a-1)(b-1)} \qquad (\omega_n = \exp(2 \pi i/n)). 
\end{align*}
We should remark that $F^{(n)}$ is unitary. The walk defined by the Fourier matrix is called the {\em Fourier walk}. Then, $A^{(m)}$ and $A^{(f)}$ are unitary, so the walks determined by them become QWs. In a similar fashion, we obtain 
\begin{cor}
\begin{align*}
\overline{\zeta} \left(A^{(s)}, T^2_N, u \right)^{-1} 
& =\exp \left[ \frac{1}{N^2} \sum^{N-1}_{k_1 =0} \sum^{N-1}_{k_2 =0} \log \left( F^{(s)} \left( \widetilde{k}_1, \widetilde{k}_{2}, u \right) \right) \right],
\\
\lim_{N \to \infty} \overline{\zeta} \left(A^{(s)}, T^2_N, u \right)^{-1} 
& = \exp \left[ \int_0^{2 \pi} \int_0^{2 \pi} \log \left( F^{(s)} \left( \theta_1, \theta_2, u \right) \right) \frac{d \theta_1}{2 \pi} \frac{d \theta_2}{2 \pi} \right],
\end{align*}
for $s \in \{m,f\}$, where
\begin{align*}
&F^{(m)} \left( w_1, w_2, u \right) 
= 1 - \frac{1+i}{2} \left( \cos w_{1} + \sin w_{1} + \cos w_{2} + \sin w_{2} \right) u 
\\
& 
\quad - \frac{1-i}{2} \left( 1 - \cos (w_{1} - w_{2}) \right) u^2 
+ \frac{1+i}{2} \left( \cos w_{1} + \sin w_{1} + \cos w_{2} + \sin w_{2} \right) u^3 - i u^4,
\\
&F^{(f)} \left( w_1, w_2, u \right) 
= 1 - \left( \cos w_{1} - \cos w_{2} \right) u 
\\
& 
\quad 
+ \frac{1-i}{2} \left( 1 - \cos (w_{1} - w_{2}) \right) u^2 
+ i \left( \cos w_{1} - \cos w_{2} \right) u^3 - i u^4.
\end{align*}
\label{kimarid2fourie}
\end{cor}
Note that $F^{(s)} \left( w_1, w_2, u \right)$ for $s \in \{m,f\}$ are the same as the corresponding results in \cite{AsanoEtAl2019}.

Finally, we consider the two-dimensional positive-support version of the Grover walk whose $4 \times 4$ coin matrix $A^{(m)}$ (M-type) or $A^{(f)}$ (F-type) is determined by 
\begin{align*}
A^{(m)} 
=
\begin{bmatrix}
0 & 1 & 1 & 1 \\ 
1 & 0 & 1 & 1 \\ 
1 & 1 & 0 & 1 \\ 
1 & 1 & 1 & 0 
\end{bmatrix} 
, \qquad 
A^{(f)} 
= 
\begin{bmatrix}
1 & 0 & 1 & 1 \\
0 & 1 & 1 & 1 \\ 
1 & 1 & 1 & 0 \\
1 & 1 & 0 & 1 
\end{bmatrix}
.
\end{align*}
Here the {\em positive support} $A^+ = [ A^+_{ab} ]$ of a real matrix $A = [ A_{ab}]$ is defined as follows:
\begin{align*}
A^+_{ab} =\left\{
\begin{array}{ll}
1 & \mbox{if $A_{ab} > 0$, } \\
0 & \mbox{if $A_{ab} \le 0$}.
\end{array}
\right.
\end{align*}
So the positive support $G^{(n),+}$ of the $n \times n$ Grover matrix $G^{(n)}$ for $n \ge 2$ is 
\begin{align*}
G^{(n),+}_{ab} =\left\{
\begin{array}{ll}
1 & \mbox{if $a \not= b$, } \\
0 & \mbox{if $a = b$}.
\end{array}
\right.
\end{align*}
We should remark that this model is neither QW nor CRW. However, we can apply our method to it. Thus we have the following results.
\begin{cor}
\begin{align*}
\overline{\zeta} \left(A^{(m)}, T^2_N, u \right)^{-1} 
& =\exp \left[ \frac{1}{N^2} \sum^{N-1}_{k_1 =0} \sum^{N-1}_{k_2 =0} \log \left( F^{(m)} \left( \widetilde{k}_1, \widetilde{k}_{2}, u \right) \right) \right],
\\
\lim_{N \to \infty} \overline{\zeta} \left(A^{(m)}, T^2_N, u \right)^{-1} 
& = \exp \left[ \int_0^{2 \pi} \int_0^{2 \pi} \log \left( F^{(m)} \left( \theta_1, \theta_2, u \right) \right) \frac{d \theta_1}{2 \pi} \frac{d \theta_2}{2 \pi} \right],
\\
\overline{\zeta} \left(A^{(f)}, T^2_N, u \right)^{-1} 
& = (1-u^2) \exp \left[ \frac{1}{N^2} \sum^{N-1}_{k_1 =0} \sum^{N-1}_{k_2 =0} \log \left( F^{(f)} \left( \widetilde{k}_1, \widetilde{k}_{2}, u \right) \right) \right],
\\
\lim_{N \to \infty} \overline{\zeta} \left(A^{(f)}, T^2_N, u \right)^{-1} 
& = (1-u^2) \exp \left[ \int_0^{2 \pi} \int_0^{2 \pi} \log \left( F^{(f)} \left( \theta_1, \theta_2, u \right) \right) \frac{d \theta_1}{2 \pi} \frac{d \theta_2}{2 \pi} \right],
\end{align*}
for $s \in \{m,f\}$, where
\begin{align*}
F^{(m)} \left( w_1, w_2, u \right) 
&= 1 - 2 \left( 1 + 2 \cos w_1 \cos w_2 \right) u^2  - 4 \left( \cos w_{1} + \cos w_{2} \right) u^3 - 3 u^4, 
\\
F^{(f)} \left( w_1, w_2, u \right) 
&= 1 - 2 \left( \cos w_{1} + \cos w_{2} \right) u + 3 u^2. 
\end{align*}
\label{cortorusGMFpsd2}
\end{cor} 
We should note that our result for F-type in Corollary \ref{cortorusGMFpsd2} corresponds to Eq. (10) in Clair \cite{Clair}. 


\section{Higher-Dimensional Case \label{sec06}} 
In this section, we consider walks on the $d$-dimensional torus $T^d_N \ (d \ge 3)$ with $2d \times 2d$ coin matrix $A^{(m)}$ (M-type) or $A^{(f)}$ (F-type). To do so, we introduce the $n$-th {\em elementary symmetric polynomial} $e_{j}^{(n)} (x_1,x_2, \ldots,x_n)$ as follows:
\begin{align*}
e_{j}^{(n)} (x_1,x_2, \ldots,x_n) = \sum_{T \subset [n], \ |T|=j} \  \prod_{t \in T} \ x_t \qquad (j=1,2, \ldots, n),
\end{align*}
where $[n] = \{1,2, \ldots,n \}$. For example, 
\begin{align*}
e_{1}^{(2)} (x_1,x_2) 
&= x_1 + x_2, \quad e_{2}^{(2)} (x_1,x_2) = x_1 x_2, 
\\
e_{1}^{(3)} (x_1,x_2,x_3) 
&= x_1 + x_2 + x_3, \quad e_{2}^{(3)} (x_1,x_2,x_3) = x_1 x_2 + x_1 x_3 + x_2 x_3, \quad 
\\
e_{3}^{(3)} (x_1,x_2,x_3) 
&= x_1 x_2 x_3. 
\end{align*}
Moreover we put
\begin{align*}
e_{j}^{(n, \cos)} (\widetilde{\kvec}) 
&=  e_{j}^{(n)} (\cos \widetilde{k}_1, \cos \widetilde{k}_2, \ldots, \cos \widetilde{k}_n), 
\\
e_{j}^{(n, \cos)} (\Theta^{(n)}) 
&=  e_{j}^{(n)} (\cos \theta_1, \cos \theta_2, \ldots, \cos \theta_n), 
\end{align*}
for $j=1,2, \ldots, n$.

A typical example is the $d$-dimensional Grover walk with $2 d \times 2 d$ coin matrix $A^{(m)}$ (M-type) or $A^{(f)}$ (F-type), where $A^{(m)}$ is the $2d \times 2d$ Grover matrix and $A^{(f)} = (I_d \otimes \sigma) A^{(m)}$. 

First we consider F-type case, since we can obtain the result on the general $d$-dimensional torus. In fact, Theorem \ref{th001} gives
\begin{cor}  
\begin{align*}
\overline{\zeta} \left(A^{(f)}, T^d_N, u \right)^{-1} 
&= (1- u^2)^{d-1} \exp \Bigg[ \frac{1}{N^d } \sum^d_{j=1} \sum^{N-1}_{ k_j =0} \log \Bigg\{ F^{(f)} \left( \widetilde{\kvec}, u \right) \Bigg\} \Bigg],
\\
\lim_{N \to \infty} \overline{\zeta} \left(A^{(f)}, T^d_N, u \right)^{-1} 
&=  (1- u^2)^{d-1} \exp \Bigg[ \int_{[0,2 \pi)^d} \log \Bigg\{ F^{(f)} \left( \Theta^{(d)}, u \right) \Bigg\} d \Theta^{(d)}_{unif} \Bigg],
\end{align*}
where
\begin{align*}
F^{(f)} \left( \wvec, u \right) 
= 1 - \frac{2}{d} \ e_{1}^{(d, \cos)} (\wvec) \  u + u^2.
\end{align*}
\label{cortorusGFgenareld}
\end{cor} 
Remark that the leading factor $(1- u^2)^{d-1}$ for $d \ge 3$ corresponds to localization of the Grover walk on $\ZM^d$ (see \cite{KomatsuKonno}, for example). Komatsu et al. \cite{KomatsuEtAl2021} obtained the same result by not our method based on the Fourier analysis but the Konno-Sato theorem (including $d=2$ case) and called ``Grover/Generalized-Zeta Correspondence".

On the other hand, as for M-type case, we do not get the result on the general $d$-dimensional torus. Indeed, Theorem \ref{th001} for $d=3$ case implies the following result. However it is not a simple form compared with the corresponding F-type.

\begin{cor}  
\begin{align*}
\overline{\zeta} \left(A^{(m)}, T^3_N, u \right)^{-1} 
&= (1- u^2) \exp \Bigg[ \frac{1}{N^3 } \sum^3_{j=1} \sum^{N-1}_{ k_j =0} \log \Bigg\{ F^{(m)} \left( \widetilde{\kvec}, u \right) \Bigg\} \Bigg],
\\
\lim_{N \to \infty} \overline{\zeta} \left(A^{(m)}, T^3_N, u \right)^{-1} 
&=  (1- u^2) \exp \Bigg[ \int_{[0,2 \pi)^3} \log \Bigg\{ F^{(m)} \left( \Theta^{(3)}, u \right) \Bigg\} d \Theta^{(3)}_{unif} \Bigg],
\end{align*}
where
\begin{align*}
F^{(m)} \left( \wvec, u \right) 
=  1 + \frac{4}{3} \ e_{1}^{(3, \cos)} (\wvec) \ u + \left( 2 + \frac{4}{3} \ e_{2}^{(3, \cos)} (\wvec) \right) \ u^2 + \frac{4}{3} \ e_{1}^{(3, \cos)} (\wvec) \ u^3 + u^4.
\end{align*}
\label{cortorusGMd3}
\end{cor} 

Another typical example is the positive-support version of the $d$-dimensional Grover walk with $2 d \times 2 d$ coin matrix $A^{(m)}$ (M-type) or $A^{(f)}$ (F-type), where $A^{(m)}$ is the positive-support of the $2d \times 2d$ Grover matrix and $A^{(f)} = (I_d \otimes \sigma) A^{(m)}$ like $d=2$ case. The situation in this example is similar to that of the previous one. So, we first deal with F-type case, since we can obtain the result for the general $d$-dimensional torus. From Theorem \ref{th001}, we have
\begin{cor}  
\begin{align*}
\overline{\zeta} \left(A^{(f)}, T^d_N, u \right)^{-1} 
&= (1- u^2)^{d-1} \exp \Bigg[ \frac{1}{N^d } \sum^d_{j=1} \sum^{N-1}_{ k_j =0} \log \Bigg\{ F^{(f)} \left( \widetilde{\kvec}, u \right) \Bigg\} \Bigg],
\\
\lim_{N \to \infty} \overline{\zeta} \left(A^{(f)}, T^d_N, u \right)^{-1} 
&=  (1- u^2)^{d-1} \exp \Bigg[ \int_{[0,2 \pi)^d} \log \Bigg\{ F^{(f)} \left( \Theta^{(d)}, u \right) \Bigg\} d \Theta^{(d)}_{unif} \Bigg],
\end{align*}
where
\begin{align*}
F^{(f)} \left( \wvec, u \right) 
= 1 - 2 \ e_{1}^{(d, \cos)} (\wvec) \ u + (2d-1) u^2.
\end{align*}
\label{cortorusGFpsgenareld}
\end{cor} 
This result is consistent with the $d$-dimensional torus $T^d_N$ case for Theorem 1.3 in Chinta et al. \cite{ChintaEtAl}. Komatsu et al. \cite{KomatsuEtAl2021} got the same result by not our method based on the Fourier analysis but the Konno-Sato theorem (including $d=2$ case) and called ``Grover(Positive Support)/Generalized-Ihara-Zeta Correspondence".

As in the case of the Grover walk, for M-type case, we do not have the result on the general $d$-dimensional torus. Actually, Theorem \ref{th001} for $d=3$ case gives the following result which is a complicated form compared with the corresponding F-type.
\begin{cor}  
\begin{align*}
\overline{\zeta} \left(A^{(m)}, T^3_N, u \right)^{-1} 
&= \exp \Bigg[ \frac{1}{N^3 } \sum^3_{j=1} \sum^{N-1}_{ k_j =0} \log \Bigg\{ F^{(m)} \left( \widetilde{\kvec}, u \right) \Bigg\} \Bigg],
\\
\lim_{N \to \infty} \overline{\zeta} \left(A^{(m)}, T^3_N, u \right)^{-1} 
&= \exp \Bigg[ \int_{[0,2 \pi)^3} \log \Bigg\{ F^{(m)} \left( \Theta^{(3)}, u \right) \Bigg\} d \Theta^{(3)}_{unif} \Bigg],
\end{align*}
where
\begin{align*}
F^{(m)} \left( \wvec, u \right) 
& = 1 - \bigg( 3 + 4 \ e_{2}^{(3, \cos)} (\wvec) \bigg) \ u^2 - 8 \bigg( e_{1}^{(3, \cos)} (\wvec) + 2 \ e_{3}^{(3, \cos)} (\wvec) \bigg) \ u^3 
\\
&
\qquad \qquad \qquad - 3 \bigg( 3 + 4 \ e_{2}^{(3, \cos)} (\wvec) \bigg) \ u^4 - 8 \ e_{1}^{(2, \cos)} (\wvec) \ u^5 - 5 u^6.
\end{align*}
\label{cortorusGMpsd3}
\end{cor} 
\par
\
\par
\
\par\noindent
{\bf Appendix A: Grover/Zeta Correspondence}
\par
\
\par\noindent
In Appendix A, we briefly review our previous work on Grover/Zeta Correspondence based on the Konno-Sato theorem in \cite{KomatsuEtAl2021}. We assume that all graphs are simple. Here we consider the Grover walk with F-type and the positive-support version of the Grover walk with F-type on a graph.

Let $G=(V(G),E(G))$ be a connected graph (without multiple edges and loops) with the set $V(G)$ of vertices and the set $E(G)$ of unoriented edges $uv$ joining two vertices $u$ and $v$. Moreover, let $n=|V(G)|$ and $m=|E(G)|$ be the number of vertices and edges of $G$, respectively. For $uv \in E(G)$, an arc $(u,v)$ is the oriented edge from $u$ to $v$. Let $D_G$ be the symmetric digraph corresponding to $G$. Set $D(G)= \{ (u,v),(v,u) \mid uv \in E(G) \} $. For $e=(u,v) \in D(G)$, set $u=o(e)$ and $v=t(e)$. Furthermore, let $e^{-1}=(v,u)$ be the {\em inverse} of $e=(u,v)$. For $v \in V(G)$, the {\em degree} $\deg {}_G \ v = \deg v = d_v $ of $v$ is the number of vertices adjacent to $v$ in $G$. If $ \deg {}_G \ v=k$ (constant) for each $v \in V(G)$, then $G$ is called {\em $k$-regular}. A {\em path $P$ of length $n$} in $G$ is a sequence $P=(e_1, \ldots ,e_n )$ of $n$ arcs such that $e_i \in D(G)$, $t( e_i )=o( e_{i+1} ) \ (1 \leq i \leq n-1)$. If $e_i =( v_{i-1} , v_i )$ for $i=1, \cdots , n$, then we write $P=(v_0, v_1, \ldots ,v_{n-1}, v_n )$. Put $ \mid P \mid =n$, $o(P)=o( e_1 )$ and $t(P)=t( e_n )$. Also, $P$ is called an {\em $(o(P),t(P))$-path}. We say that a path $P=( e_1 , \ldots , e_n )$ has a {\em backtracking} if $ e^{-1}_{i+1} =e_i $ for some $i \ (1 \leq i \leq n-1)$. A $(v, w)$-path is called a {\em $v$-cycle} (or {\em $v$-closed path}) if $v=w$. Let $B^r$ be the cycle obtained by going $r$ times around a cycle $B$. Such a cycle is called a {\em multiple} of $B$. A cycle $C$ is {\em reduced} if both $C$ and $C^2 $ have no backtracking. The {\em Ihara zeta function} of a graph $G$ is a function of a complex variable $u$ with $|u|$ sufficiently small, defined by 
\begin{align*}
{\bf Z} (G, u)= \exp \left( \sum^{\infty}_{r=1} \frac{N_r}{r} u^r \right), 
\end{align*}
where $N_r$ is the number of reduced cycles of length $r$ in $G$. Let $G$ be a connected graph with $n$ vertices $v_1, \ldots ,v_n $. The {\em adjacency matrix} ${\bf A}= {\bf A} (G)=(a_{ij} )$ is the square matrix such that $a_{ij} =1$ if $v_i$ and $v_j$ are adjacent, and $a_{ij} =0$ otherwise. The following result was obtained by Ihara \cite{Ihara} and Bass \cite{Bass}.
\begin{theorem}[Ihara \cite{Ihara}, Bass \cite{Bass}]
Let $G$ be a connected graph. Then the reciprocal of the Ihara zeta function of $G$ is given by 
\begin{align*}
{\bf Z} (G,u )^{-1} =(1- u^2 )^{\gamma-1} 
\det \left( {\bf I} -u {\bf A} (G)+ u^2 ( {\bf D} - {\bf I} ) \right). 
\end{align*}
Here $\gamma$ is the Betti number of $G$, and ${\bf D} =( d_{ij} )$ is the diagonal matrix with $d_{ii} = \deg v_i$ and $d_{ij} =0, i \neq j$, where $V(G)= \{ v_1 , \ldots , v_n \}$. 
\end{theorem}

Let $G=(V(G),E(G))$ be a connected graph with $n$ vertices and $ x_0 \in V(G)$ a fixed vertex. Then the {\em generalized Ihara zeta function} $\zeta (G, u)$ of $G$ is defined by 
\begin{align*}
\zeta (G, u)= \exp \left( \sum^{\infty}_{r=1} \frac{N^0_r}{r} u^r \right), 
\end{align*}
where $N^0_r$ is the number of reduced $x_0$-cycles of length $r$ in $G$. A graph $G$ is called {\em vertex-transitive} if there exists an automorphism $ \phi $ of the automorphism group $Aut(G)$ of $G$ such that $ \phi (u)=v$ for each $u,v \in V(G)$. Note that if $G$ is a vertex-transitive graph with $n$ vertices, then 
\begin{align*}
\zeta (G, u)= {\bf Z} (G,u)^{1/n}. 
\end{align*}
Moreover, the {\em Laplacian} of $G$ is given by 
\begin{align*}
\Delta {}_{n} = \Delta (G) = {\bf D} - {\bf A} (G). 
\end{align*}

A formula for the generalized Ihara zeta function of a vertex-transitive graph is given in Chinta et al. \cite{ChintaEtAl} as follows.

\begin{theorem}[Chinta et al. \cite{ChintaEtAl}]
Let $G$ be a  vertex-transitive $(q+1)$-regular graph with spectral measure $\mu {}_{\Delta }$ for the Laplacian $\Delta $. 
Then 
\begin{align*}
\zeta (G, u)^{-1} 
=(1-u^2 )^{(q-1)/2} \exp \left[ \int \log (1-(q+1- \lambda )u+q u^2 ) d \mu {}_{\Delta } ( \lambda )  \right]. 
\end{align*}
\label{ThChinta}
\end{theorem}

Let $G$ be a connected graph with $n$ vertices and $m$ edges. Put $V(G)= \{ v_1 , \ldots , v_n \} $ and $d_j = d_{v_j} = \deg v_j , \ j=1, \ldots , n$. Then the {\em Grover matrix} ${\bf U} ={\bf U} (G)=( U_{ef} )_{e,f \in D(G)} $ of $G$ is defined by 
\begin{align*}
U_{ef} =\left\{
\begin{array}{ll}
2/d_{t(f)} (=2/d_{o(e)} ) & \mbox{if $t(f)=o(e)$ and $f \neq e^{-1} $, } \\
2/d_{t(f)} -1 & \mbox{if $f= e^{-1} $, } \\
0 & \mbox{otherwise. }
\end{array}
\right. 
\end{align*}
The discrete-time quantum walk with the matrix ${\bf U} $ as a time evolution matrix is the Grover walk with F-type on $G$. Let $G$ be a connected graph with $n$ vertices and $m$ edges. Then the $n \times n$ matrix ${\bf P}_{n} = {\bf P} (G)=( P_{uv} )_{u,v \in V(G)}$ is given by
\begin{align*}
P_{uv} =\left\{
\begin{array}{ll}
1/( \deg {}_G \ u)  & \mbox{if $(u,v) \in D(G)$, } \\
0 & \mbox{otherwise.}
\end{array}
\right.
\end{align*}
Note that the matrix ${\bf P} (G)$ is the transition probability matrix of the simple random walk on $G$. We introduce the {\em positive support} ${\bf F}^+ =( F^+_{ij} )$ of a real matrix ${\bf F} =( F_{ij} )$ as follows.
\begin{align*}
F^+_{ij} =\left\{
\begin{array}{ll}
1 & \mbox{if $F_{ij} >0$, } \\
0 & \mbox{otherwise}.
\end{array}
\right.
\end{align*}
Ren et al. \cite{RenEtAl} showed that the Perron-Frobenius operator (or edge matrix) of a graph is the positive support $({}^{\rm{T}}{\bf U})^+ $ of the transpose of its Grover matrix ${\bf U} $, i.e., 
\begin{align*}
{\bf Z} (G,u)^{-1} = \det \left( {\bf I}_{2m} -u( {}^{\rm{T}}{\bf U})^+ \right)= \det \left( {\bf I}_{2m} -u {\bf U}^+ \right). 
\end{align*}
The Ihara zeta function of a graph $G$ is just a zeta function on the positive support of the Grover matrix of $G$. That is, the Ihara zeta function corresponds to the positive-support version of the Grover walk (defined by the positive support of the Grover matrix ${\bf U}^+$) with F-type on $G$.

Now we propose a new zeta function of a graph. Let $G$ be a connected graph with $m$ edges. Then we define a zeta function $ \overline{{\bf Z}} (G, u)$ of $G$ satisfying 
\begin{align*}
\overline{{\bf Z}} (G, u)^{-1} = \det ( {\bf I}_{2m} -u {\bf U} ).    
\end{align*}
In other words, the new zeta function corresponds to the Grover walk (defined by the Grover matrix ${\bf U}$) with F-type on $G$.

In this setting, Konno and Sato \cite{KonnoSato} presented the following result which is called the {\em Konno-Sato theorem}. 

\begin{theorem}[Konno and Sato \cite{KonnoSato}]
Let $G$ be a connected vertex-transitive $(q+1)$-regular graph with $n$ vertices and $m$ edges. Then  
\begin{align*}  
\overline{{\bf Z}} (G, u)^{-1} 
= \det ( {\bf I}_{2m} - u {\bf U} )
&=(1-u^2)^{m-n} \det \left( (1+u^2) {\bf I}_{n} -2u {\bf P}_{n} \right)
\\
&=(1-u^2)^{m- n} \det \left( (1-2u+u^2) {\bf I}_{n} + \frac{2u}{q+1} {\bf \Delta}_{n} \right),
\\
{\bf Z} (G,u)^{-1}
= \det ( {\bf I}_{2m} - u {\bf U}^+ )
&=(1-u^2)^{m- n} \det \left( (1+qu^2) {\bf I}_{n} -(q+1)u {\bf P}_{n} \right) 
\\
&=(1-u^2)^{m-n} \det \left( (1-(q+1)u+qu^2) {\bf I}_{n} +u {\bf \Delta}_{n} \right). 
\end{align*}
\label{KS} 
\end{theorem}
Here we give a weight functions $w: D(G) \times D(G) \longrightarrow \mathbb{C} $ as follows. 
\begin{align*}
w (e,f) =\left\{
\begin{array}{ll}
2/ \deg t(e)  & \mbox{if $t(e)=o(f)$ and $f \neq e^{-1} $, } \\
2/ \deg t(e) -1 & \mbox{if $f= e^{-1} $, } \\
0 & \mbox{otherwise. }
\end{array}
\right.
\end{align*}
For a cycle $C=( e_1, e_2 , \ldots , e_r )$, put
\begin{align*}
w(C)=w(e_1 , e_2 ) \cdots w( e_{r-1} , e_r) w( e_r , e_1 ). 
\end{align*}

We define a generalized zeta function with respect to the Grover matrix of a graph. Let $G=(V(G),E(G))$ be a connected graph and $ x_0 \in V(G)$ a fixed vertex. Then the {\em generalized zeta function} $\overline{\zeta} {}_G (u)$ of $G$ is defined by 
\begin{align*}
\overline{\zeta} (G, u) = \exp \left( \sum^{\infty}_{r=1} \frac{N^0_r }{r} u^r \right), 
\end{align*}
where 
\begin{align*}
N^0_r = \sum \left\{ w(C) \mid C: \ an \ x_0 \hbox{-} cycle \ of \ length \ r \ in \ G \right\}. 
\end{align*}
We should remark that if $G$ is a vertex-transitive graph with $n$ vertices, then 
\begin{align*}
\overline{\zeta} (G, u)
= \overline{{\bf Z}} (G,u)^{1/n} . 
\end{align*}

Then the following result for a series of finite vertex-transitive $(q+1)$-regular graphs was given in our previous work \cite{KomatsuEtAl2021}, which is called {\em Grover/Zeta Correspondence}.

\begin{theorem}[Grover/Zeta Correspondence \cite{KomatsuEtAl2021}]  
Let $\{ G_m \}^{\infty}_{m=1} $ be a series of finite vertex-transitive $(q+1)$-regular graphs with $\lim_{m \to \infty} |V(G_m)|= \infty.$ Then
\begin{align*}   
\lim_{m \to \infty} \overline{\zeta} (G_m, u)^{-1} 
&= (1-u^2 )^{(q-1)/2} \exp \left[ \int \log \left\{ (1+u^2)-2u \lambda \right\} d \mu_{P} ( \lambda ) \right]
\\   
&= (1-u^2 )^{(q-1)/2} \exp \left[ \int \log \left\{ (1-2u+u^2) + \frac{2u}{q+1} \lambda \right\} d \mu_{\Delta} ( \lambda ) \right], 
\\
\lim_{m \to \infty} \zeta (G_m, u)^{-1} 
&= (1-u^2 )^{(q-1)/2} \exp \left[ \int \log \left\{ (1+q u^2)-(q+1)u \lambda \right\} d \mu_P ( \lambda ) \right]
\\    
&= (1-u^2 )^{(q-1)/2} \exp \left[ \int \log \left\{ (1+q u^2)-((q+1)- \lambda ) u \right\} d \mu_{\Delta } ( \lambda ) \right],   
\end{align*}
where $d \mu_P ( \lambda )$ and $d \mu_{\Delta } ( \lambda )$ are the spectral measures for the transition operator ${\bf P}$ and the Laplacian $\Delta$. 
\label{tiktok01}
\end{theorem} 
We should note that the fourth formula in Theorem \ref{tiktok01} is nothing but Theorem 1.3 in Chinta et al. \cite{ChintaEtAl} (see also Theorem \ref{ThChinta} in Appendix A).

Next the following result for the generalized zeta function and the generalized Ihara zeta function of the {\em $d$-dimensional torus} $T^d_N$ was shown in our previous work \cite{KomatsuEtAl2021}, which is also called {\em Grover/Zeta Correspondence} ({\em $T^d_N$ case}). Note that $|E( T^d_N )|=d N^d$ and $T^d_N$ is a vertex-transitive $2d$-regular graph.

\begin{theorem}[Grover/Zeta Correspondence ($T^d_N$ case) \cite{KomatsuEtAl2021}]    
Let $T^d_N \ (d \geq 2)$ be the $d$-dimensional torus with $N^d$ vertices. Then we have
\begin{align*} 
\lim_{N \to \infty} \overline{\zeta} (T^d_N ,u)^{-1} 
&= (1- u^2 )^{d-1} 
\exp \left[ \int_{[0,2 \pi)^d}   
\log \left\{ (1+u^2 )- \frac{2u}{d} \sum^d_{j=1} \cos \theta_j \right\} d \Theta^{(d)}_{unif} \right],  
\\
\lim_{N \to \infty} {\zeta} (T^d_N ,u)^{-1} 
&= (1- u^2 )^{d-1} 
\exp \left[ \int_{[0,2 \pi)^d} 
\log \left\{ (1+(2d-1) u^2 )-2u \sum^d_{j=1} \cos \theta_j \right\} d \Theta^{(d)}_{unif} \right],
\end{align*}
where $\int_{[0,2 \pi)^d}$ is the $d$-th multiple integral and $d \Theta^{(d)}_{unif}$ is the uniform measure on $[0, 2 \pi )^d $.  
\end{theorem} 
The first result on $\lim_{N \to \infty} \overline{\zeta} (T^d_N ,u)^{-1}$ is the same as that on $\lim_{N \to \infty} \overline{\zeta} (A^{(f)}, T^d_N ,u)^{-1}$ for the $d$-dimensional Grover walk with F-type in Corollary \ref{cortorusGFgenareld} (Section \ref{sec06}). The second result on $\lim_{N \to \infty} \zeta (T^d_N ,u)^{-1}$ is the same as that on $\lim_{N \to \infty} \zeta (A^{(f)}, T^d_N ,u)^{-1}$ for the positive-support version of the $d$-dimensional Grover walk with F-type in Corollary \ref{cortorusGFpsgenareld} (Section \ref{sec06}).

Specially, in the case of $d=2$, the following result can be derived. 
\begin{cor}     
Let $T^2_N $ be the $2$-dimensional torus with $N^2$ vertices. Then we have
\begin{align*} 
\lim_{N \to \infty} \overline{\zeta} (T^2_N ,u)^{-1} 
&= (1- u^2) \exp \left[ \int^{2 \pi}_{0} \int^{2 \pi}_{0}  
\log \left\{ (1+u^2 )-u \sum^d_{j=1} \cos \theta_j \right\} \frac{d \theta_1}{2 \pi } \frac{d \theta_2}{2 \pi } \right],  
\\
\lim_{N \to \infty} {\zeta} (T^2_N ,u)^{-1} 
&= (1- u^2) \exp \left[ \int^{2 \pi}_{0} \int^{2 \pi}_{0}  
\log \left\{ (1+3 u^2 )-2u \sum^2_{j=1} \cos \theta_j \right\} \frac{d \theta_1}{2 \pi } \frac{d \theta_2}{2 \pi } \right].   
\end{align*}
\end{cor} 
The first result on $\lim_{N \to \infty} \overline{\zeta} (T^2_N ,u)^{-1}$ is equivalent to that on $\lim_{N \to \infty} \overline{\zeta} (A^{(f)}, T^2_N ,u)^{-1}$ for the two-dimensional Grover walk with F-type in Corollary \ref{kimarid2grover} (Section \ref{sec05}). The second result on $\lim_{N \to \infty} \zeta (T^2_N ,u)^{-1}$ is equivalent to that on $\lim_{N \to \infty} \zeta (A^{(f)}, T^2_N ,u)^{-1}$ for the positive-support version of the two-dimensional Grover walk with F-type in Corollary \ref{cortorusGMFpsd2} (Section \ref{sec05}). 



\end{document}